\documentclass{article}

\usepackage{PRIMEarxiv}

\usepackage[utf8]{inputenc} 
\usepackage[T1]{fontenc}    
\usepackage{hyperref}       
\usepackage{url}            
\usepackage{booktabs}       
\usepackage{amsfonts,amsmath}       
\usepackage{nicefrac}       
\usepackage{microtype}      
\usepackage{lipsum}
\usepackage{fancyhdr}       
\usepackage{graphicx}       
\usepackage{natbib}
\usepackage{algorithm}
\usepackage{algpseudocode}

\title{Robust variable selection for spatial point processes observed with noise}

\author{
  Dominik Sturm$^{1,2,3,4}$ \\
  \texttt{sturm@mpi-cbg.de} \\
   \And
  Ivo F. Sbalzarini$^{1,2,3,4,5}$\thanks{Corresponding author.}  \\
  \texttt{sbalzarini@mpi-cbg.de} \\
  \AND
  \\
  $^1$Dresden University of Technology, Faculty of Computer Science, Dresden, Germany \\
  $^2$Max Planck Institute of Molecular Cell Biology and Genetics, Dresden, Germany \\
  $^3$Center for Systems Biology Dresden, Dresden, Germany \\
  $^4$Center for Scalable Data Analytics and Artificial Intelligence ScaDS.AI, Dresden/Leipzig, Germany \\
  $^5$Cluster of Excellence Physics of Life, Technical University Dresden, Dresden, Germany \\
}

\begin{document}
\maketitle

\begin{abstract}
We propose a method for variable selection in the intensity function of spatial point processes that combines sparsity-promoting estimation with noise-robust model selection. As high-resolution spatial data becomes increasingly available through remote sensing and automated image analysis, identifying spatial covariates that influence the localization of events is crucial to understand the underlying mechanism. However, results from automated acquisition techniques are often noisy, for example due to measurement uncertainties or detection errors, which leads to spurious displacements and missed events. We study the impact of such noise on sparse point-process estimation across different models, including Poisson and Thomas processes. To improve noise robustness, we propose to use stability selection based on point-process subsampling and to incorporate a non-convex best-subset penalty to enhance model-selection performance. In extensive simulations, we demonstrate that such an approach reliably recovers true covariates under diverse noise scenarios and improves both selection accuracy and stability. We then apply the proposed method to a forestry data set, analyzing the distribution of trees in relation to elevation and soil nutrients in a tropical rain forest. This shows the practical utility of the method, which provides a systematic framework for robust variable selection in spatial point-process models under noise, without requiring additional knowledge of the process.
\end{abstract}


\section{Introduction}\label{sec:introduction}

Spatial data, described through the locations of points or events, is ubiquitous in applications. It is found in ecology \citep{renner_equivalence_2013, renner_point_2015}, forestry \citep{waagepetersen_two-step_2009}, epidemiology \citep{zimmerman_estimating_2008, diggle_statistical_2013}, cell biology \citep{helmuth_beyond_2010, parra_methods_2021, summers_spatial_2022}, and telecommunication \citep{li_statistical_2015}. Such data can be statistically modeled by spatial point processes. A spatial point-process model represents probabilities of observations as a random subset $X\subseteq W$, where $W$ is an observation window of interest, i.e., the domain of the data. A fundamental quantity in spatial point processes is the expected number of events over $W$, which is described by the {\em intensity} of the process as the first moment of the distribution over $X$.

In practical applications, the intensity is often unknown, as the point process is observed through data, i.e., realizations of a point process.
A common task then is to estimate the intensity at a location $u\in W$ from an observed point pattern with respect to some covariates $z(u)\in\mathbb{R}^p$, $p\geq 1$. If $p$ is large ($p\gg 1$) it becomes necessary---both for interpretability and computational efficiency---to use automatic variable-selection procedures \citep{hastie_statistical_2015}. The goal of variable selection is to identify a small set of covariates that are collectively sufficient to explain the observed spatial distribution of points.

The problem of variable selection has been addressed from different viewpoints. Fitting regularized maximum-likelihood estimators has been proposed for Poisson and clustering processes \citep{thurman_variable_2014,thurman_regularized_2014,choiruddin_convex_2018}, as well as for Gibbs point processes \citep{yue_variable_2015, ba_inference_2023}.
It has also been shown how such regularization techniques can be used to detect the correlation structure of highly multivariate point processes \citep{rajala_detecting_2018, choiruddin_regularized_2020}.
\citep{thurman_regularized_2014, choiruddin_convex_2018, ba_inference_2023} have established asymptotic results for the resulting estimators in an increasing domain for certain regularizations and processes.
An alternative approach solved auxiliary tasks for variable-importance measures \citep{spychala_variable_2024}.
These works have established the possibility and advantages of variable selection in point-process modeling.

Variable selection in spatial point processes is not only possible, but practically viable.
Experimental results have demonstrated that adaptive regularization methods, such as the adaptive Lasso \citep{zou_adaptive_2006}, are particularly effective in recovering the sparse support of the model \citep{choiruddin_convex_2018,ba_inference_2023,coeurjolly_regularization_2023}. However, Lasso-type penalties represent convex relaxations of the true variable-selection problem.

Here, we go beyond convex relaxations by directly considering an $L_0$ penalty for variable selection. Although this renders the resulting optimization problem non-convex, the $L_0$ penalty promises (in theory) faster convergence in the risk than the Lasso, without requiring assumptions on the design matrix \citep{bach_learning_2024}. Realizing this theoretical advantage, however, is hard, since algorithms for $L_0$ optimization with guarantees have exponential time complexity. We therefore leverage proximal operators to compute approximate solutions without theoretical guarantees on global optimality. We show in extensive numerical experiments that the sparse local minima thus identified are sufficient to recover the true support of the model in many cases, while providing great performance improvements over the Lasso.

In addition to the performance of variable selection, also its robustness to noise is important in practical applications.
While information criteria have been shown to reliably recover important variables in spatial point processes \citep{choiruddin_convex_2018, choiruddin_information_2021, ba_inference_2023}, they do not account for uncertainty in the data.
It is therefore somewhat surprising that the robustness of point-process variable selection under noise has received relatively little attention so far \citep{guttorp_what_2023}.
Nevertheless, we think this is an interesting topic, since noise on the observed point patterns can influence the performance of variable selection in nontrivial ways.
Noise can occur by design, for example, in presence-only analysis in ecology \citep{renner_point_2015}, where events are modeled only if they are detected and not necessarily everywhere they are present.
But it can also occur through automated data analysis, for example, in remote sensing \citep{gillespie_deep_2024}, biomedical image analysis \citep{kuronen_point_2021,parra_methods_2021}, and automated geocoding \citep{zimmerman_estimating_2008,briz-redon_dealing_2024}.

In point processes, noise comes in two flavors: point localization errors due to measurement uncertainties and misdetected events due to detection errors.
Previous works either focused on explicitly modeling noise and correcting the estimates \citep{lund_models_2000,kuronen_point_2021,zimmerman_estimating_2008,briz-redon_dealing_2024} or on
cross validation \citep{yue_variable_2015, rajala_detecting_2018, choiruddin_regularized_2020}.
While this improves the accuracy of the estimates, it is inconsequential for the robustness of the estimator.
Moreover, cross validation becomes difficult if repeated experiments are not available or the number of samples is small, as correlations can lead to overcomplete models.
Noise correction requires a statistical noise model, which might not be available in applications where the data-acquisition process is not well understood.

Instead of modeling and correcting for the noise, we show how the proposed $L_0$ penalty can be used in a robust statistic to directly improve the noise robustness of the estimator without assumptions about the noise process.
This avoids the tendency of the Lasso to overfit the data under noise \citep{werner_loss-guided_2023}.
We propose to use stability selection \citep{meinshausen_stability_2010,shah_variable_2013}, which has shown to increase variable selection performance in $L_0$-regularized problems \citep{maddu_stability_2022}, as well as selection stability compared to the Lasso \citep{nogueira_stability_2018}. Stability selection is based on subsampling the data. For point processes, subsampling can be done by $p$-thinning. We believe this to be a promising avenue, since it has been shown that subsampling-based estimators are effective at finding the optimal bandwidth of kernel density estimates of point processes \citep{cronie_cross-validation-based_2024}. We therefore derive estimating equations for stability selection over point processes. This allows controlling the per-family error rate (PFER) of the estimator under noise \citep{meinshausen_stability_2010} and is easily integrated into existing algorithms.

We validate the proposed algorithm on synthetic data of simulated point processes with varying noise levels and different noise types. The experiments suggest that the proposed stability selection algorithm is able to recover the true covariates under noise while being robust to overfitting. We compare the performance of the proposed method for $L_0$ and $L_1$ penalties to existing methods, including the adaptive Lasso and various information criteria.
While stability selection introduces a computational overhead that scales linearly with the number of subsamples,
we show that it greatly improves variable-selection performance and robustness of the estimator, in particular with the sparser $L_0$ penalty, while achieving error control.
To show the practical utility of our method, we apply it to a real-world forestry data set, where we analyze tree occurrences in relation to elevation and soil nutrients. The models identified by the proposed method are consistent with other methods, but tend to contain fewer variables and do not require knowledge of the underlying process.
We therefore believe that the proposed method is a useful addition to the toolbox of spatial point-process modeling.

\section{Methods}\label{sec:methods}

We review the formulation of spatial point processes considered here and introduce the notation. Then, we present the methods for sparse intensity estimation and model selection.

\subsection{Spatial Point Processes and Estimating Function Inference}\label{sec:methods:pp}

A spatial point process $X$ is a random process on $W\subseteq \mathbb{R}^d$, $d\geq 1$, whose realizations $X=\{u_i\}_{i=1}^n$ are a locally finite subset of $W$. We denote $N(B)=|X \cap B|$ as the number of points in some region $B\subseteq W$ and say $X$ is locally finite if $N(B)<\infty$ for all $B\subseteq W$. As the distribution of $X$ is in many cases difficult to express, either through counting measures or void probabilities, statistical inference usually revolves around the characterization of point processes through joint intensity functions \citep{moller_modern_2007, baddeley_spatial_2007}. Given a set of points $u_1,\dots, u_n$ the $n$-th order joint intensity $\rho^{(n)}(u_1,\dots, u_n)\Delta u_1,\dots, \Delta u_n$ can be interpreted as the probability of finding one point in each of the infinitesimal regions $\Delta u$. More formally, we can define $\rho^{(n)}$ as the $n$-th order joint intensity through
\begin{multline}\label{eq:methods:pp:intensity}
  \mathbb{E}\!\left\{\sum\nolimits^{\neq}_{u_1,\dots,u_n\in X}
  \mathbf{1}(u_1\in B_1,\dots, u_n\in B_n)\right\} \\
  = \int_{B_1}\dots\int_{B_n}\mathbf{1}(u_1\in B_1,\dots, u_n\in B_n)
  \rho^{(n)}(u_1,\dots, u_n)\,\mathrm{d}u_1\dots\mathrm{d}u_n.
\end{multline}
As such, $\rho^{(1)}$ and $\rho^{(2)}$ characterize the first and second moments of the distribution. We abbreviate $\rho(u)$ for the first-order intensity $\rho^{(1)}(u)$ and define the pair correlation
\begin{align}
  g(u,v) = \left\{
    \begin{array}{ll}
      0  & \text{ if } \rho(u)\rho(v) = 0\, , \\
      \displaystyle\frac{\rho^{(2)}(u,v)}{\rho(u)\rho(v)} \ & \text{ otherwise\,.}
    \end{array}
    \right.
  \end{align}
  Using Palm conditioning, the product $\rho(u)g(u,v)$ can be interpreted as the intensity of observing a point at $u$ given that $v\in X$ \citep{coeurjolly_tutorial_2017}. A value of $g(u,v)>1$ increases the likelihood of observing an additional point at $u$ and therefore indicates {\em spatial attraction}. Conversely, $g(u,v)<1$ decreases the probability of finding an additional point in some neighborhood and thus indicates {\em spatial inhibition}.

In the following, we consider processes that are second-order stationary reweighted, i.e., they only vary in their first moment, and their covariance is stationary \citep{diggle_statistical_2013}. Further restricting to isotropic interactions, this implies that the pair correlation simplifies to only depend on the distance $r$ between two points, i.e., $g(u,v)=g(r)$. Many such processes can be modeled using a separable set of parameters $\theta=(\beta^\top, \alpha^\top)\in\mathbb{R}^{p_l+q}$. Here, $\beta=(\omega, \beta_1,\dots, \beta_p)^\top\in\mathbb{R}^{p_l}$ specifies the parameters of the intensity, which is given as a log-linear model $\rho(u;\beta)=\omega\exp\left\{\beta_{1:p}^\top z(u)\right\}$ depending on a set of spatial covariates $z(u)$ and $\omega>0$ setting the scale. The parameters $\alpha\in\mathbb{R}^q$ model spatial interactions via the pair-correlation function $g(r;\alpha)$.

Since the likelihood is intractable for processes other than the Poisson process, the use of composite likelihood approaches has become popular for parametric estimation \citep{moller_modern_2007}. Campbell's formula \citep{baddeley_spatial_2007}
  can be used to specify a system of unbiased estimating equations \citep{lavancier_adaptive_2021}. Given the above assumptions, we can estimate the mean using the Poisson score function
  \begin{equation}\label{eq:methods:pp:first_order_estimator}
    e(\beta) := \sum_{u\in X}\frac{\nabla_\beta \rho(u;\beta)}{\rho(u;\beta)} - \int_W\nabla_\beta \rho(u;\beta)\,\mathrm{d}u=0\, .
  \end{equation}
  This can be seen as the limit of composite likelihoods of Bernoulli random trials over partitions $c_i\in W$ as $\prod_{c_i}{(\rho(c_i)|c_i|)}^{N_i}{(1-\rho(c_i)|c_i|)}^{1-N_i}$, for $N_i=\mathbf{1}(N(c_i)>0)$, and it provides an unbiased estimating equation even if the underlying model is not Poisson \citep{moller_modern_2007, moller_recent_2017}\footnote{This can be seen from Campbell's formula, choosing $f(u)=\nabla_\beta\log\rho(u;\beta)$ as the test function.}. Similarly, a second-order estimating equation can be constructed \citep{waagepetersen_estimating_2007, lavancier_adaptive_2021} as
  \begin{equation}\label{eq:methods:pp:second_order_estimator}
    0 = \sum_{(u,v)\in X}^{\neq}\frac{\nabla_\alpha g(u,v;\alpha)}{g(u,v;\alpha)} - \int_W\int_W\rho(u;\beta)\rho(v;\beta)\nabla_\alpha g(u,v;\alpha)\,\mathrm{d}u\mathrm{d}v\, .
  \end{equation}
  Estimating equations of this family also include the second-order composite likelihood introduced by \citep{guan_composite_2006} and Palm likelihoods by conditioning \citep{prokesova_two-step_2017}.

  Alternatively, one can use estimators based on the radially symmetric pair-correlation function $g(r)$, or the $K$-function, for the number of events observed at distance $r$. If $g(r)$ or $K(r)$ are known, non-parametric models $\hat{g}(r)$ or $\hat{K}(r)$ can be used to estimate $\alpha$ using minimum-contrast estimation \citep{moller_modern_2007,waagepetersen_two-step_2009}. Like the estimator in Eq.~\eqref{eq:methods:pp:second_order_estimator}, the non-parametric models $\hat{g}(r)$ and $\hat{K}(r)$ also depend on an initial estimate of the intensity $\rho(u;\beta)$, since they describe the ``normalized'' second-order properties of the process~\citep{baddeley_non_2000}. Using the $K$-function, the objective function for the estimation of $\alpha$ can be written as
  \begin{equation}\label{eq:methods:pp:minimum_contrast}
    \hat{\alpha} =\arg\min_{\alpha\in\mathbb{R}^q} \int_{r_{\min}}^{r_{\max}} \left[K(r;\alpha)^b - \hat{K}(r)^b\right]^2 \mathrm{d}r\, .
  \end{equation}
  This minimizes the distance between the estimated and observed $K$-functions over some range $[r_{\min}, r_{\max}]$, which has to be chosen small enough to capture local variations. The exponent $b$ controls the variance of the estimator; it is chosen empirically \citep{diggle_statistical_2013}.

  In this paper, we consider two types of point processes in order to assess how the underlying correlation structure affects variable-selection performance. Specifically, we consider the Poisson point process, in which events are uncorrelated, and the Thomas point process, which induces spatial attraction. In the latter, event clusters can arise either due to an inhomogeneous intensity function or from attractive interactions between points. This makes variable selection more challenging.

  In a Poisson point process over $W$, points are independently distributed according to an intensity function $\rho(u)$. We say that a process $X$ is Poisson over $W$ with intensity $\rho(u)$ if the number of points $N(W)\sim\mathbf{Poisson}\left(\int_W\rho(u)\,\mathrm{d}u\right)$, where $\mathbf{Poisson}(\cdot)$ is the Poisson distribution,
  and each point $u\in X$ is {\it i.i.d.}~with probability density $p(u)=\rho(u)/\int_W\rho(u)\, \mathrm{d}u$ \citep{moller_statistical_2003}. The Poisson point process is one of the rare cases for which the distribution is explicitly known. Because of this, point-process densities are often specified w.r.t.~the unit-rate Poisson process, that is, a Poisson process with $\rho(u)=1$. Doing so, the probability density of the Poisson process with intensity $\rho(u)$ is
  \begin{equation}\label{eq:methods:pp:poisson_density}
    f(X) = \exp\left\{|W|-\int_W\rho(u)\, \mathrm{d}u\right\}\prod_{u\in X}\rho(u) \, .
  \end{equation}
  From this, we see that the estimating equation in Eq.~\eqref{eq:methods:pp:first_order_estimator} is the score of a Poisson point process log-likelihood.

  The second model we consider is the Thomas point process, which corresponds to a shot-noise Cox process~\citep{moller_statistical_2003}. Cox processes are often referred to as {\em doubly stochastic}~\citep{diggle_statistical_2013}, since their intensity function is itself a realization of a random process. The intensity function of a Cox process \textit{driven by} a non-negative random field $\Gamma(u)$ with finite variance \citep{moller_statistical_2003}, is given as:
  \begin{equation}
    \rho(u)=\mathbb{E}\{\Gamma(u)\}\, ,
  \end{equation}
  and its pair-correlation function is:
  \begin{equation}
    g(u,v)=\frac{\mathbb{E}\{\Gamma(u)\Gamma(v)\}}{\rho(u)\rho(v)}\, .
  \end{equation}

  To construct a Thomas point process, consider a random field $\Gamma(u)$ defined by {\em parent points} $Y$ sampled from a Poisson point process with intensity $\kappa>0$.
  Conditional on each $v\in Y$, consider the Poisson point processes $X_v$ with intensity
  \begin{equation}
    \rho_v(u)=\omega\exp\{\beta_{1:p}^\top z(u)\}\mathbf{Normal}(u;v,\sigma^2\mathbb{I}_d)/\kappa\, ,
  \end{equation}
  where $\mathbf{Normal}(\cdot)$ is the $d$-dimensional normal distribution, and $\omega>0$ controls the number of {\em daughter points}. The Thomas point process is then given by the superposition of all daughter-point patterns $\cup_{v\in Y}X_v$. It has intensity
  \begin{equation}
    \rho(u)=\omega\exp\{\beta_{1:p}^\top z(u)\}\, ,
  \end{equation}
  which is of log-linear form and can be estimated using Eq.~\eqref{eq:methods:pp:first_order_estimator} \citep{moller_modern_2007,choiruddin_convex_2018}. From this definition, it is apparent that the daughter points around each parent $v\in Y$ admit clustering. In this way, all shot-noise Cox processes can be seen as generalized cluster processes~\citep{prokesova_two-step_2017}. For the Thomas process this is captured by the pair-correlation function
  \begin{equation}
    g(u,v) = 1 + \frac{\exp\{-\big\Vert u-v\big\Vert_2^2 / (4\sigma^2)\}}{4\pi\kappa\sigma^2}\, ,
  \end{equation}
  which is attractive ($g(r)\geq 1$) for all $r>0$~\citep{moller_statistical_2003}.

  \subsection{Sparse Intensity Estimation}

  Parametric estimation of point-process models often revolves around identifying how spatial covariates $z(u)$ influence the expected number of points in some region of interest. If the dimensionality of $z(u)$ rapidly increases with $p\gg 1$, the question of sparse variable selection naturally arises. For point processes, we can adapt the Poisson likelihood in Eq.~\eqref{eq:methods:pp:first_order_estimator} to regularize the amount of selected variables similar to the Lasso:
  \begin{equation}\label{eq:methods:sparsity:objective}
    \log\ell(\beta) = \sum_{u\in X}\log \rho(u;\beta) - \int_W\log\rho(u;\beta)\, \mathrm{d}u - \sum_{i=1}^p \lambda_i h(\beta_i)\, .
  \end{equation}
  Here, $h$ denotes a penalty function to be chosen. Since previous studies argued for the use of adaptive penalties, we here use the adaptive Lasso, which has been found to perform best in for both intensity \citep{choiruddin_convex_2018} and conditional intensity \citep{ba_inference_2023} estimation tasks.
  Following \citep{zou_adaptive_2006}, we set the adaptive penalty to $\lambda_i=\lambda / |\hat\beta_i|$, where $\hat\beta$ is the unpenalized maximizer of Eq.~\eqref{eq:methods:sparsity:objective}.
  This provides a convex relaxation of the actual variable-selection problem.

  Sparse variable selection penalizes the number of nonzero coefficients in $\beta$. This defines an $L_0$ problem with penalty function $\Vert\beta\Vert_0:=\sum_{i=1}^p \mathbf{1}(\beta_i \neq 0)$. This, however, constitutes a best-subset selection problem, which is NP-hard to solve exactly \citep{bach_learning_2024}.
  Moreover, the $L_0$ penalty function is not differentiable, hampering the use of gradient-based optimizers. Nevertheless, it can be approximately solved using proximal gradient descent (PGD), which provides a general framework for non-differentiable penalties \citep{yang_fast_2020}.
  The proximal operator of a penalty function $h$ is defined as $\mathrm{prox}_h(x):=\arg\min_{v\in \mathbb{R}^p}\!\left[ h(v) + \frac{1}{2}\big\Vert v-x\big\Vert_2^2\right] $.
  This amounts to a generalized projection operator~\citep{parikh_proximal_2014}. PGD computes a series of parameter updates by gradient descent over the negative Poisson likelihood followed by applying the proximal operator of the penalty. For the $L_0$-constrained problem with penalty weight $\lambda$, this becomes
  \begin{align}
    \beta^{t+1} &= \mathrm{prox}_{\lambda\gamma\Vert\cdot\Vert_0}\left(\beta^{t}+\gamma e(\beta)\right)\\
    &=\arg\min_{v\in \mathbb{R}^p}\left[ \lambda\Vert v\big\Vert_0 + \frac{1}{2\gamma}\big\Vert v-\left(\beta^{t}+\gamma e(\beta)\right)\big\Vert_2^2 \right] \\
    &=T_{\sqrt{2\lambda\gamma}}\left(\beta^{t}+\gamma e(\beta)\right)\, .
  \end{align}
  Here, $\gamma$ is the step size of the gradient descent, and $T_\xi$ is the hard-thresholding operator
  \begin{equation}
    T_\xi(\beta) =
    \begin{cases}
      \beta & |\beta| > \xi\\
      0 & \mathrm{else}\, .
    \end{cases}
  \end{equation}
  The same algorithm, with the correspondingly changed proximal operator, also applies to  Lasso and elastic net penalties and their adaptive variants. Then, $T_\xi$ becomes the soft-thresholding or scaled soft-thresholding operator,
  respectively, resulting in the classic ISTA algorithm~\citep{hastie_statistical_2015}. This is summarized in Table~\ref{tab:proximal_operators}.

  \begin{table}
    \caption{Penalty functions $h$ and their corresponding proximal operators, where $\lambda$ is the penalty weight and $\gamma$ the gradient-descent step size (see Algorithm~\ref{algorithm}). All proximal operators are evaluated element-wise.}
    \centering\footnotesize
    \begin{tabular*}{\linewidth}{@{\extracolsep{\fill}} c c p{3.5cm}}
      Penalty $h$  & $\mathrm{prox}_{h}$ & Name \\\hline\\
      $\big\Vert\beta\big\Vert_0$  & $T_{\gamma\lambda}(\beta)=
      \begin{cases}
        \beta & \beta^2 > 2\gamma\lambda\\
        0 & \mathrm{else}
      \end{cases}$ & best-subset selection \hspace{1cm} hard thresholding \\\\
      $\big\Vert\beta\big\Vert_1$ & $S_{\gamma\lambda}(\beta)=
      \begin{cases}
        \beta - \gamma\lambda & \beta > \gamma\lambda\\
        \beta + \gamma\lambda & \beta < -\gamma\lambda\\
        0 & \mathrm{else}
      \end{cases}$ & Lasso \hspace{2cm} soft thresholding  \\
    \end{tabular*}
    \label{tab:proximal_operators}
  \end{table}

  We obtain a fast algorithm for first-order PGD by computing an adaptive step size $\gamma\in\mathbb{R}_+$ using the Barzilai--Borwein (BB) method~\citep{barzilai_two-point_1988}. This computes the step size from a scalar approximation of the Hessian by solving the least-squares problem
  \begin{equation}
    \hat\gamma = \arg\min_{\gamma\in\mathbb{R}_+} \big\Vert \Delta\beta - \gamma\Delta g\big\Vert_2^2 = |\Delta\beta^\top\Delta g| / \big\Vert\Delta g\big\Vert_2^2\, ,
  \end{equation}
  where $\Delta\beta = \beta^t - \beta^{t-1}$ and $\Delta g= e(\beta^t) -e(\beta^{t-1})$
  for two subsequent iterations $t-1$ and $t$. We observed empirically for the $L_1$ problem that PGD with such an adaptive step size outperforms PGD with fixed step size and accelerated versions both in speed and stability~\citep{beck_fast_2009, hastie_statistical_2015}. In some of the cases presented below, however, we observed that BB step sizes can induce oscillations when using the $L_0$ penalty, where a model term oscillates between being in the support and being set to zero. This is due to the dependence of the hard thresholding on the step size $\gamma$ close to the inclusion boundary. Therefore, we use BB-adaptive steps only for the $L_1$ penalty. For the $L_0$ penalty, we use a fixed step size of $\gamma=10^{-3}$. We find that this performs comparably to accelerated PGD~\citep{yang_fast_2020}.

  The algorithm stops upon convergence. We detect convergence if the likelihood does not increase over 1000 subsequent iterations or if the relative change in $\beta$ is below $\epsilon=10^{-4}$ between two subsequent iterations.

  \subsection{Model Selection}\label{sec:selection}

  The penalization weight $\lambda$ controls the number of selected variables and hence the complexity of the resulting model. Choosing $\lambda$ is crucial. This is typically done using information criteria, such as the Akaike (AIC) or Bayesian information criteria (BIC) \citep{choiruddin_convex_2018, choiruddin_information_2021, choiruddin_adaptive_2023}.
  However, information criteria underperform for misspecified likelihoods, e.g., when estimating non-Poisson models using Eq.~\eqref{eq:methods:pp:first_order_estimator}, because they ignore the correlation structure \citep{choiruddin_information_2021}. The model may then include additional covariates to compensate for unmodeled clustering in the data.

  Correcting for unmodeled correlations, composite information criteria for the estimating equation in Eq. \eqref{eq:methods:pp:first_order_estimator} have been derived \citep{choiruddin_information_2021, ba_inference_2023}. This includes the composite BIC \citep{choiruddin_information_2021}
  \begin{equation}\label{eq:methods:model_selection:cbic}
    \mathbf{cBIC}(\hat\beta) = -2\log\ell(\beta) + \mathbf{df}(\rho_\beta)\log N(W)\, ,
  \end{equation}
  where $\mathbf{df}(\rho_\beta)=\mathrm{tr}(S^{-1}\Sigma)$ are the effective degrees of freedom with sensitivity matrix $S$ and asymptotic variance--covariance matrix $\Sigma$. While $S$ can be estimated from the Hessian of $\log\ell(\beta)$, estimating $\Sigma$ requires additional knowledge about the pair-correlation function $g(r)$ of the process.  \citep{choiruddin_information_2021} proposed the simplified estimator
  \begin{equation}\label{eq:methods:model_selection:cbic:df}
    \mathbf{df}(\rho_\beta) = k + \mathrm{tr}(S^{-1}T_2)\, ,
  \end{equation}
  with $k>0$ the number of non-zero entries in $\hat\beta$ and
  \begin{equation}\label{eq:methods:model_selection:cbic:covariance}
    T_2 = \int_W\!\int_W \!z(u)z(v)^\top\! \rho(u;\hat{\beta})\rho(v;\hat{\beta})(g(u,v) -1)\,\mathrm{d}u\mathrm{d}v \, .
  \end{equation}
  This requires estimating $g(u,v)$ using a valid parametric model \citep{choiruddin_information_2021}, which implies additional assumptions on the underlying process. If the underlying process is Poisson, i.e., if $g(u,v)=1$, the term $T_2$ vanishes and $\mathbf{df}(\rho_\beta) = k$ reduces to the standard BIC \citep{choiruddin_information_2021}.

  Another generalization of the BIC to penalized likelihoods has been proposed with the extended regularized information criterion (ERIC) \citep{hui_tuning_2015}. The composite ERIC (cERIC) has been used to estimate Lasso-regularized conditional-intensity models of Gibbs point processes \citep{ba_inference_2023}. It directly includes knowledge of the regularization parameter $\lambda$ to weight the degrees of freedom:
  \begin{equation}
    \mathbf{cERIC}(\hat\beta) = -2\log\ell(\beta) + \mathbf{df}(\rho_\beta)\log\!\left(\!\frac{N(W)}{\lambda}\!\right).
  \end{equation}
  Composite information criteria like cBIC and cERIC can outperform the standard BIC by favoring models with higher penalization \citep{choiruddin_information_2021, ba_inference_2023}.

  Using composite information criteria on real data, however, poses two challenges: First, neither the pair-correlation function nor the intensity function of the underlying point process are usually known. Second, while composite information criteria correct for correlations in the data, they do not account for noise.
  We therefore propose an alternative model selection procedure based on stability selection. This does not require modeling the second-order moments, and it improves noise robustness by combining knowledge over several subsamplings of the data \citep{meinshausen_stability_2010}. In the context of sparse regression, stability selection can be understood as a bootstrap  estimate of the inclusion probability of a covariate $\beta_j$ in the support of the model $\mathcal{S}$. As the support depends on the choice of $\lambda\geq 0$, we estimate the support $\mathcal{S}^\lambda(Z_i)$ over $K$
  subsamples $Z_i,\, i=1,\ldots , K$,
  of the observed process $X$ over the regularization path $\lambda\in[\lambda_{\min}, \lambda_{\max}]=:\Lambda$. The stability measure is then defined as \citep{meinshausen_stability_2010}:
  \begin{equation}\label{eq:methods:selection:stability}
    \Pi_j^\lambda:=\mathbb{P}\{ j\in \mathcal{S}^\lambda\}\approx\frac{1}{K}\sum_{i=1}^K\mathbf{1}(j\in \mathcal{S}^\lambda(Z_i))\, ,
  \end{equation}
  which converges by the law of large numbers given a sufficient bootstrap size $K$. In our experiments, we find that $K=50$ was sufficient. Larger $K$ did not significantly improve the results. The decision to select a covariate $\beta_j$ is taken by thresholding the inclusion probability $\mathbb{P}\{ j\in \mathcal{S}^\lambda\}$, i.e., $\beta_j$ is selected if and only if $\Pi_j^\lambda\geq\pi_{\mathrm{th}}$. The threshold $\pi_{\mathrm{th}}$ defines the required stability for a term to be included in the model. It is usually chosen between 0.7 and 0.9.

  A distinctive advantage of stability selection over (composite) information criteria and cross validation is the possibility for error control. \citep{meinshausen_stability_2010} derived an upper bound on the per-family error rate (PFER), which is the expected number of falsely selected variables, as a function of the selection threshold $\pi_{\mathrm{th}}$ and the expected number of selected terms over the regularization path $\Lambda$, $q_\Lambda=\mathbb{E}\{ |\cup_{\lambda\in\Lambda}\mathcal{S}^\lambda|\}$. Under an exchangeability assumption on the inclusion probabilities of noise variables, this bound is
  \begin{equation}\label{eq:methods:selection:stability:error_control}
    \mathrm{PFER}\leq \frac{1}{2\pi_{\mathrm{th}} - 1}\frac{q_\Lambda^2}{p}\, .
  \end{equation}
  Following \citep{meinshausen_stability_2010}, and unless otherwise stated, we choose $\lambda_{\max}$ such that $|\mathcal{S}^{\lambda_{\max}}(Z_i)|=0$ and the $\lambda_{\min}$ with $q_\Lambda=\sqrt{0.8p}$. At $\pi_{\mathrm{th}}=0.9$ this guarantees that $\mathrm{PFER}\leq 1$. While the exchangeability assumption might not be satisfied in practice, for example if predictors are correlated, the empirical PFER in our experiments below never exceeds the bound.
  The tightness of the bound could potentially be improved using complimentary-pairs stability selection (CPSS) \citep{shah_variable_2013}. For the experiments below, this was not necessary, but we nevertheless describe how the approach presented here can be extended to CPSS. \citep{bodinier_automated_2023} proposed model-based automatic calibration of the stability-selection hyperparameters for selecting the final model from a feasible set obtained by error control.
  We find that this procedure sometimes improves the $F_1$ score, albeit at the cost of decreasing the True Positive Rate (TPR). Since this means that we might miss important covariates, we do not use automatic hyperparameter calibration here, although it can naturally be combined with our approach. Instead, and to maintain methodological simplicity, we use Eq.~\eqref{eq:methods:selection:stability:error_control} for error control.

  We use error control according to Eq.~\eqref{eq:methods:selection:stability:error_control} only to identify the support of the model $\mathcal{S}$ with stability selection over $K=50$ subsamples $Z_i$.
  The coefficient values are then estimated on the entire data $X$ with fixed model support.
  When creating the bootstrap samples $Z_i$, we distinguish between replicated experiments and single observations. For replicated experiments $\mathcal{X}=\{Z_i\}_{i=1}^N$, we draw subsets from $\mathcal{X}$ with replacement \citep{hastie_elements_2009}. If there is only a single observation, the bootstrap samples are obtained by subsampling the point pattern $X$. Similar to \citep{cronie_cross-validation-based_2024}, who used repeated subsampling of a point process to derive a point-process learning objective, we argue that subsampling $X$ should be done by independent thinning with retention probability $p_{\mathrm{thin}}:W\rightarrow [0,1]$. For this subsampling process, analytical moment expressions are available \citep{moller_statistical_2003}.
  We define an independent thinning $Z_i$ of a point process $X$ with retention probability $p_{\mathrm{thin}}(u)$ as a new point process in which each point $u \in X$ is retained independently with probability $p_{\mathrm{thin}}(u)$. This results in $Z_i=\{u:u\in X, m_u=1\}$ with $m_u\sim\mathbf{Bernoulli}(p_{\mathrm{thin}}(u))$. The intensity function of the thinned point process $Z$ is $\rho_{Z}(u)=p_{\mathrm{thin}}(u) \rho(u)$, where $\rho(u)$ is the intensity function of the original point process \citep{moller_statistical_2003}. Motivated by classic stability selection and bootstrapping, we choose $p_{\mathrm{thin}}=0.5$ unless mentioned otherwise. This also readily extends to CPSS to allow for additional error control. There, one would estimate the parameters over both $Z_i$ and $X\setminus Z_i$ and evaluate the stability for all $2K$ subsamples using Eq.~\eqref{eq:methods:selection:stability}. This could be further extended to loss-guided stability selection \citep{werner_loss-guided_2023}. In that case, the set of coefficients would be chosen to minimize a given loss function, for which loss functions based on innovation measures, as used in point-process learning \citep{cronie_cross-validation-based_2024}, 
  might be well suited.

  \begin{algorithm}[t]
    \caption{Compute the stability path $\{\Pi^{\lambda_i}\}_{i=1}^M$ for a penalty function $h$ from Table~\ref{tab:proximal_operators} using proximal gradient descent (PGD) with warm starts.}\label{algorithm}
    \begin{algorithmic}[1]
      \For{$k \in \{1, \dots, K\}$}
      \State $X_k \leftarrow \mathrm{subsample}(X, p_{\mathrm{thin}})$\Comment{multinomial or $p$-thinning}
      \State $\hat{\beta}_{\lambda_{0}} \leftarrow 0$ \Comment{initialize with $0$ for $\lambda_0=\lambda_{\max}$}
      \State $\gamma^0\leftarrow10^{-4}$ \textbf{if} $h=L_1$ \textbf{else} $10^{-3}$
      \For{$\lambda_i \in \{\lambda_{1}, \dots, \lambda_{M} : \lambda_i > \lambda_{i+1}\}$}
      \Comment{solve Eq.~\eqref{eq:methods:sparsity:objective}} 
      \State $\beta_{\lambda_i}^0 \leftarrow \hat{\beta}_{\lambda_{i-1}}$
      \For{$t \in \{1, \dots, \text{max\_iter}\}$ \textbf{or} convergence}\Comment{perform PGD}
      \If{$t>1$ \textbf{and} adaptive\_step}
      \State $\Delta \beta \leftarrow \beta_{\lambda_i}^{t-1} - \beta_{\lambda_i}^{t-2}$
      \State $\Delta g \leftarrow \nabla(-\log\ell(\beta_{\lambda_i}^{t-1}; X_k)) - \nabla(-\log\ell(\beta_{\lambda_i}^{t-2}; X_k))$
      \State $\gamma^t \leftarrow |{\Delta \beta}^\top\Delta g|/\big\Vert \Delta g\big\Vert_2^2$
      \EndIf
      \State $\beta_{\lambda_i}^t \leftarrow \mathrm{prox}_{\gamma^t\lambda_ih}(\beta_{\lambda_i}^{t-1} - \gamma^t\nabla(-\log\ell(\beta_{\lambda_i}^{t-1}; X_k)))$
      \EndFor
      \State $\hat{\beta}_{\lambda_i} \leftarrow \beta_{\lambda_i}^{\text{max\_iter}}$
      \EndFor
      \EndFor
    \end{algorithmic}
  \end{algorithm}

  Due to the known distributional properties of the thinned process, we can define estimating equations for the subsamplings $Z_i$ in stability selection:
  \begin{align}\label{eq:methods:selection:stability:estimating_equations}
    e_i(\beta) &:= \sum_{z\in Z_i}\nabla_\beta\log\rho(z)-p_{\mathrm{thin}}\int_W\!\nabla_\beta\rho(u)\,\mathrm{d}u\, .
  \end{align}
  This follows from Poisson-likelihood inference over $\rho_Z$. Therefore, each $Z_i$ can be used to obtain a bootstrap estimate of the underlying intensity function $\rho(u)$. The likelihood can be approximated using the Berman--Turner device \citep{berman_approximating_1992} as implemented in the \texttt{spatstat} package \cite{baddeley_spatial_2016}:
  \begin{equation}
    \log\ell_i(\beta) \approx \sum_{j=1}^M v_j\left( y_j\log\rho(u_j)-p_{\mathrm{thin}}\rho(u_j)\right)\, ,
  \end{equation}
  which resembles the likelihood of weighted Poisson regression with response $y_j=v_j^{-1}\mathbf{1}(u_j\in Z_i)$ and quadrature weight $v_j$.
  The Berman--Turner device enables the use of software for generalized linear models (GLM) and has also been extended to conditional intensity estimation \citep{baddeley_practical_2000}. Here, we numerically solve the estimating equations using \texttt{pytorch} \citep{paszke_pytorch_2019} by implementing the PGD algorithm and discretizing the integral in the estimating equations using midpoint quadrature.
   This provides the flexibility for using different penalties, as well as the possibility for tensorization and GPU acceleration. The pseudo-code for our implementation is given in Algorithm~\ref{algorithm}. It uses warm starts to accelerate the computation of the regularization path \citep{hastie_statistical_2015}. We empirically find this algorithm to be particularly effective at converging to \emph{sparse} local minima for the non-convex $L_0$ penalty.

  \section{Results}\label{sec:results}

  We empirically evaluate the proposed estimation procedure and compare it with model selection based on information criteria for data containing different types and levels of noise. We use simulated data with known ground truth to quantify the performance of the methods. As a baseline, we use a Poisson process with uncorrelated events. Then, we consider a Thomas process with correlated events as an example of a process with spatial attraction. We compare the variable-selection accuracy of the adaptive $L_0$ and $L_1$ penalties in conjunction with selection strategies based on information criteria and stability selection. Since estimating the intensity function of a Thomas process using the Poisson likelihood constitutes a misspecified model, we also evaluate the composite cBIC and cERIC in comparison with stability selection. Following the simulation benchmarks, we apply the proposed method to estimating the distribution of trees in a tropical rainforest. There, we aim to identify a sparse set of covariates, such as soil nutrients or topography, that influence tree distribution. This shall demonstrate the practical applicability of the proposed method.

  \subsection{Simulation benchmarks}\label{sec:results:simulation_study}

We quantify the accuracy of the estimator using synthetic data. For this, we use the covariate data available from the Barro Colorado Island (BCI) research plot in Panama \citep{condit_tropical_1998,hubbell_light-gap_1999}, which contains measurements of elevation, elevation gradients, and soil nutrients for a total of 15 covariates. We standardize all covariates and interpolate them to a common grid of size $201\times 101$, which is the standard grid size for this dataset in \texttt{spatstat}. We consider the observation window $W=[0,250]\times[0,125]$ for the Poisson process and an erosion by $4\sigma$ for the Thomas process to avoid edge effects. This is the same simulation setup as used in previous works \cite{choiruddin_convex_2018, choiruddin_information_2021, ba_inference_2023}, and it allows assessing variable selection performance under realistic covariates without needing to model the covariates by Gaussian processes. We consider both Poisson (in the following indicated by \textbf{P}) and Thomas (in the following indicated by \textbf{T}) point processes, where elevation and elevation gradients are used as the two true covariates. The intercept $\omega$ of the log-linear model is chosen to achieve a desired number of points in the observation window $W$. Specifically, we simulate $\mathbb{E}N(W)=50, 100, \dots, 250$ to investigate performance over varying sample sizes. For the Poisson process, we choose moderate effect sizes $\beta_{1:2}=(1, 0.5)^\top$, as larger effect sizes generally improve estimator performance. For the Thomas process, we consider $\kappa=4\times 10^{-3}$ and scale parameter $\sigma=1.5$, which results in a strong positive correlation between points and is adjusted for size \citep{choiruddin_information_2021}. The coefficients of the covariates are set to $\beta_{1:2}=(2, 0.75)^\top$ following previous works \citep{choiruddin_convex_2018, choiruddin_information_2021}. Therefore, one covariate has a strong effect, the other a moderate effect. In all cases we use the $201\times 101$ grid nodes as quadrature points and compute regularization paths $\Lambda = [10^{-4}, 5\times 10^{2}]$ for the Poisson process and $\Lambda = [10^{-3}, 10^{3}]$ for the Thomas process, each with 35 log-equidistant points to ensure sufficient coverage of the parameter space.

\begin{figure}
  \centering
  \includegraphics[width=\linewidth]{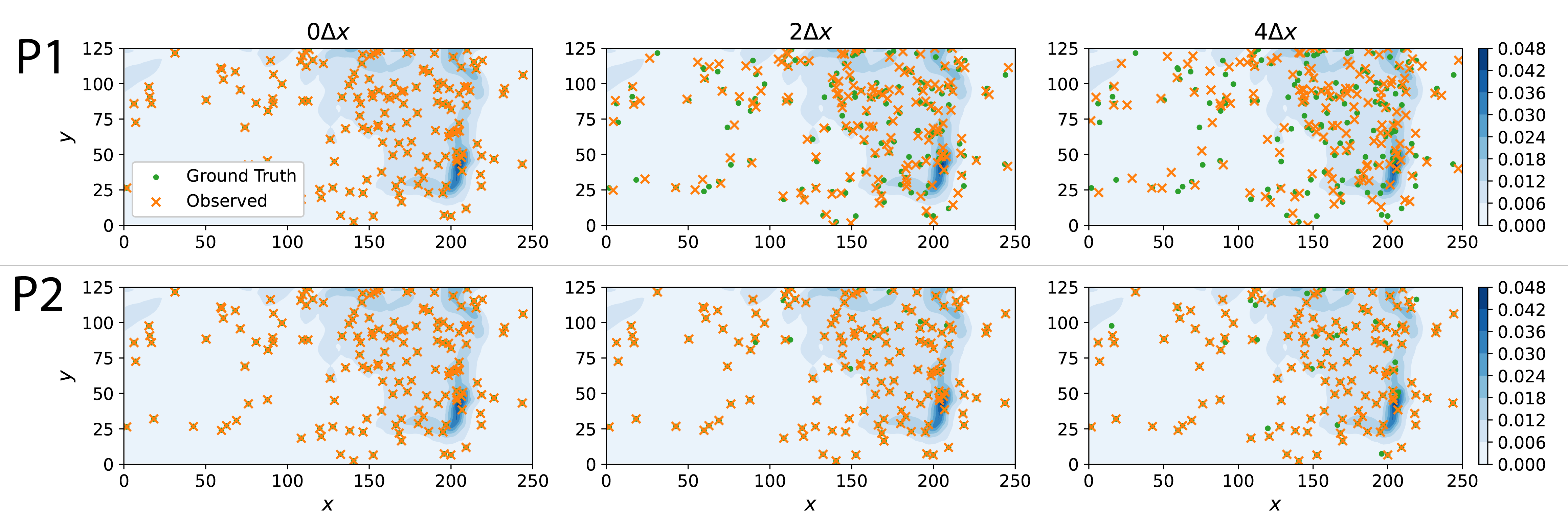}
  \caption{Illustration of the simulation setup and the noise types considered here for a Poisson point process with $\mathbb{E}N(W)=150$ and parameters $\beta=(1, 0.5)^\top$. The top row shows samples with different levels (from left to right: $c=0,2,4$) of localization uncertainty (scenario \textbf{P1}). The bottom row shows the same with detection uncertainty (scenario \textbf{P2}). Green points indicate the true simulated point locations; orange crosses show the locations observed with noise. The blue shades visualize the intensity function of the underlying point process (color bar).}
  \label{fig:results:experimental_setup}
\end{figure}

Since we are particularly interested in the robustness of the variable-selection schemes against noise in the data, we artificially corrupt the data with two different types of noise common in practical applications:
\begin{description}
  \item[Localization Uncertainty (P1/T1):] Given $\mathbb{E}N(W)$, the process (Poisson or Thomas) is simulated to obtain a sample $X$. Localization uncertainty is modeled by adding random displacements to the sampled points in $X$. We use Gaussian displacements to model measurement errors and change each point's location $u_i$ to $\tilde u_i=u_i + \epsilon$, $\epsilon\sim \mathbf{Normal}(0, \delta^2\mathbb{I}_d)$. The standard deviation $\delta=c\Delta x$ depends on the grid spacing $\Delta x$ with $c=0,1,\dots,4$ setting the dimensionless noise magnitude.
  \item[Detection Uncertainty (P2/T2):] Given $\mathbb{E}N(W)$, the process (Poisson or Thomas) is simulated to obtain a sample $X$. Detection uncertainty is modeled by missing events using distance-dependent thinning. For each point $u_i\in X$ we draw a random cutoff distance $r_i\sim |\mathbf{Normal}(0, \delta^2\mathbb{I}_d)|$ with $\delta=c\Delta x$. The point $u_i$ is retained if and only if no previously retained point lies inside the ball $B(u_i, r_i)$. The noise magnitude is set by $c=0,1,\dots,4$. 
\end{description}
Both types of noise can lead to biased estimates and affect the performance of variable-selection methods. This is particularly relevant in automated data acquisition, where location uncertainty can stem from inaccuracies in the detection process and points might be missed, e.g., due to occlusion in dense regions. Examples of noise realizations for $c=0,2,4$ are shown in Fig.~\ref{fig:results:experimental_setup} for a simulated Poisson process.

\begin{figure}
  \centering
  \includegraphics[width=0.7\linewidth,trim={0 0.7cm 0 0},clip]{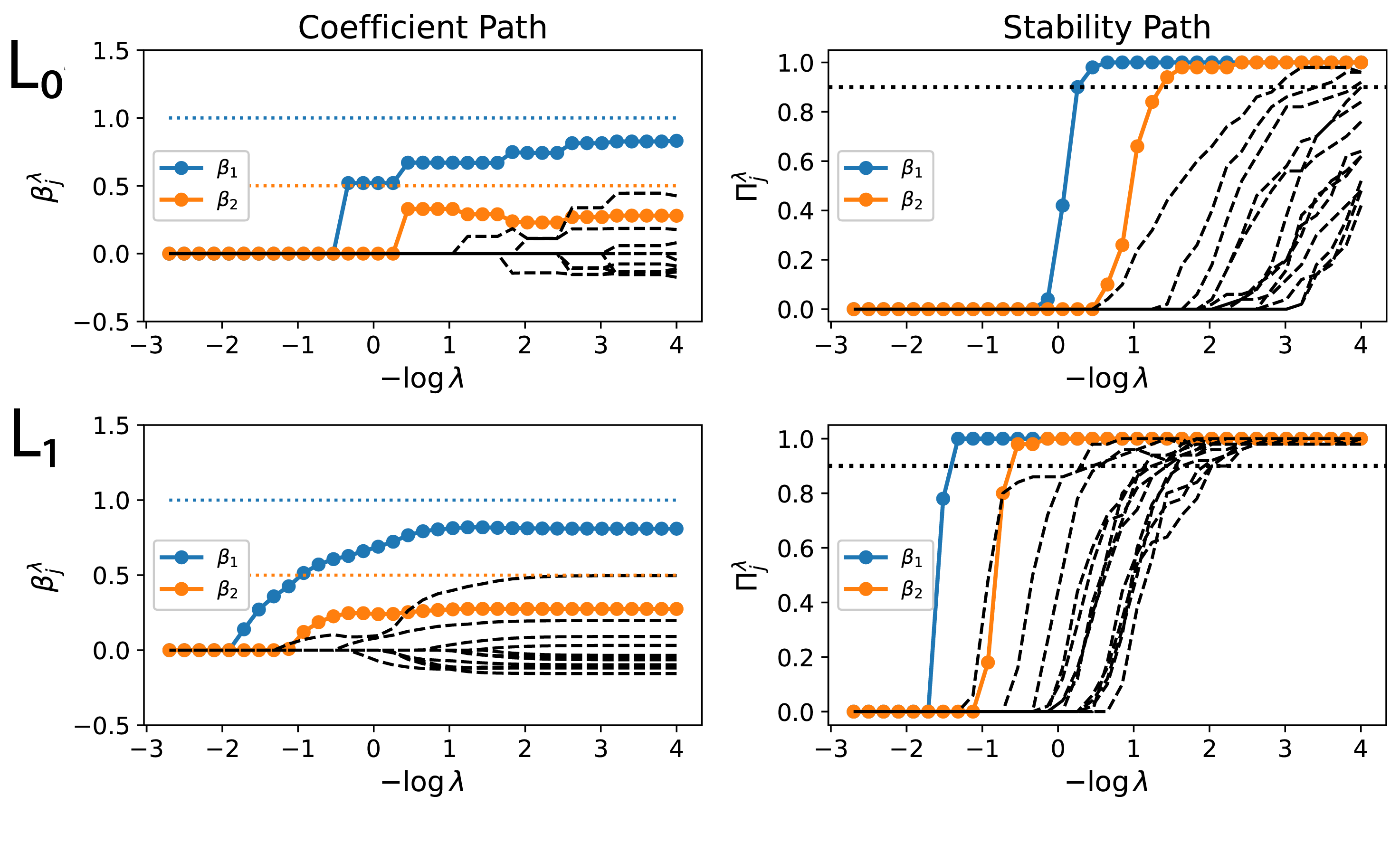}
  \caption{Regularization paths ($\lambda\in[10^{-4}, 5\times 10^{2}]$) for a Poisson process with parameters $\beta=(1, 0.5)^\top$ observed with localization uncertainty (scenario \textbf{P1}, $\mathbb{E}N(W)=200$, $c=4$). The penalty ($L_0$, $L_1$) is indicated by the row labels on the left. The left panels show the coefficient paths $\beta_j^\lambda$ with the coefficients of the true covariates
    as symbol lines (ground truth values indicated by dotted lines) and noise covariates as dashed lines.
    The right panels show the corresponding stability paths $\Pi_j^\lambda$
  with the dotted horizontal line indicating the threshold $\pi_{\mathrm{th}}=0.9$ corresponding to $\mathrm{PFER}\leq 1$.}
  \label{fig:results:simulation:poisson:paths}
\end{figure}

Figure~\ref{fig:results:simulation:poisson:paths} shows the regularization paths when using adaptive $L_0$ and $L_1$ penalties for a Poisson process with localization uncertainty (scenario \textbf{P1}) and $\mathbb{E}N(W)=200$, $c=4$. Here, the adaptive $L_0$ penalty achieves better separation between true covariates (solid colored lines with symbols) and noise covariates (dashed black lines) than the adaptive $L_1$ penalty. Notably, as seen from the coefficient paths in the left panels, the $L_1$ penalty assigns a nonzero coefficient to a noise covariate together with $\beta_2$, whereas for the $L_0$ penalty noise covariates only show up for low $\lambda$. The right panels show the stability paths $\Pi_j^\lambda$ for both penalties with the selection threshold $\pi_{\mathrm{th}}=0.9$ indicated by a dotted line. The adaptive $L_0$ penalty yields lower stability scores for noise covariates than the adaptive $L_1$ penalty. As a result, noise variables are selected only at low $\lambda$, leaving a wider range of $\lambda$ in which the true covariates are correctly identified. This means that the $L_0$-regularized estimator is more robust to noise in the data than the $L_1$ estimator.

To systematically quantify the influence of different penalties and selection criteria on the variable-selection performance under noise, we repeat the experiment for different sample sizes $\mathbb{E}N(W)$ and noise levels $c$ with 100 independent repetitions each. We measure the performance of variable selection using the True Positive Rate (TPR), False Positive Rate (FPR), Positive Predictive Value (PPV), the $F_1$ score, and the feature-selection stability $\Phi_S$ \citep{nogueira_stability_2018}. The TPR is the fraction of correctly selected covariates over all true covariates. The FPR is the fraction of selected noise covariates over all noise covariates. The PPV is the fraction of correctly selected covariates over all selected covariates. The $F_1$ score is the harmonic mean of the TPR and PPV and is a standard metric in machine learning. The feature-selection stability $\Phi_S$ was proposed by \citep{nogueira_stability_2018} and quantifies the variance of the selection indicator random variable, i.e., how reliably a feature is selected. It decreases with increasing indicator variance as:
\begin{equation}\label{eq:methods:selection:stability_metric}
  \Phi_S = 1 - \frac{\frac{1}{p}\sum_{f=1}^p s_f^2}{\frac{\overline{k}}{p}\left(1-\frac{\overline{k}}{p}\right)}\, .
\end{equation}
Here, $s_f^2=\frac{M}{M-1}\hat p_f(1-\hat p_f)$ is the empirical sample variance for selecting the $f^\text{th}$ feature with selection probability $\hat p_f$ over $M$ repetitions (here $M=100$). The denominator is the expected sample variance under random selection, with $\overline{k}$ the average number of selected features over all selections. Like all other performance metrics considered here, $\Phi_S$ is between $0$ and $1$ (for large $M$), where $0$ indicates no stability and $1$ indicates perfect stability.

\begin{figure}
  \centering
  \includegraphics[width=\linewidth,trim={0 0 0.5cm 0},clip]{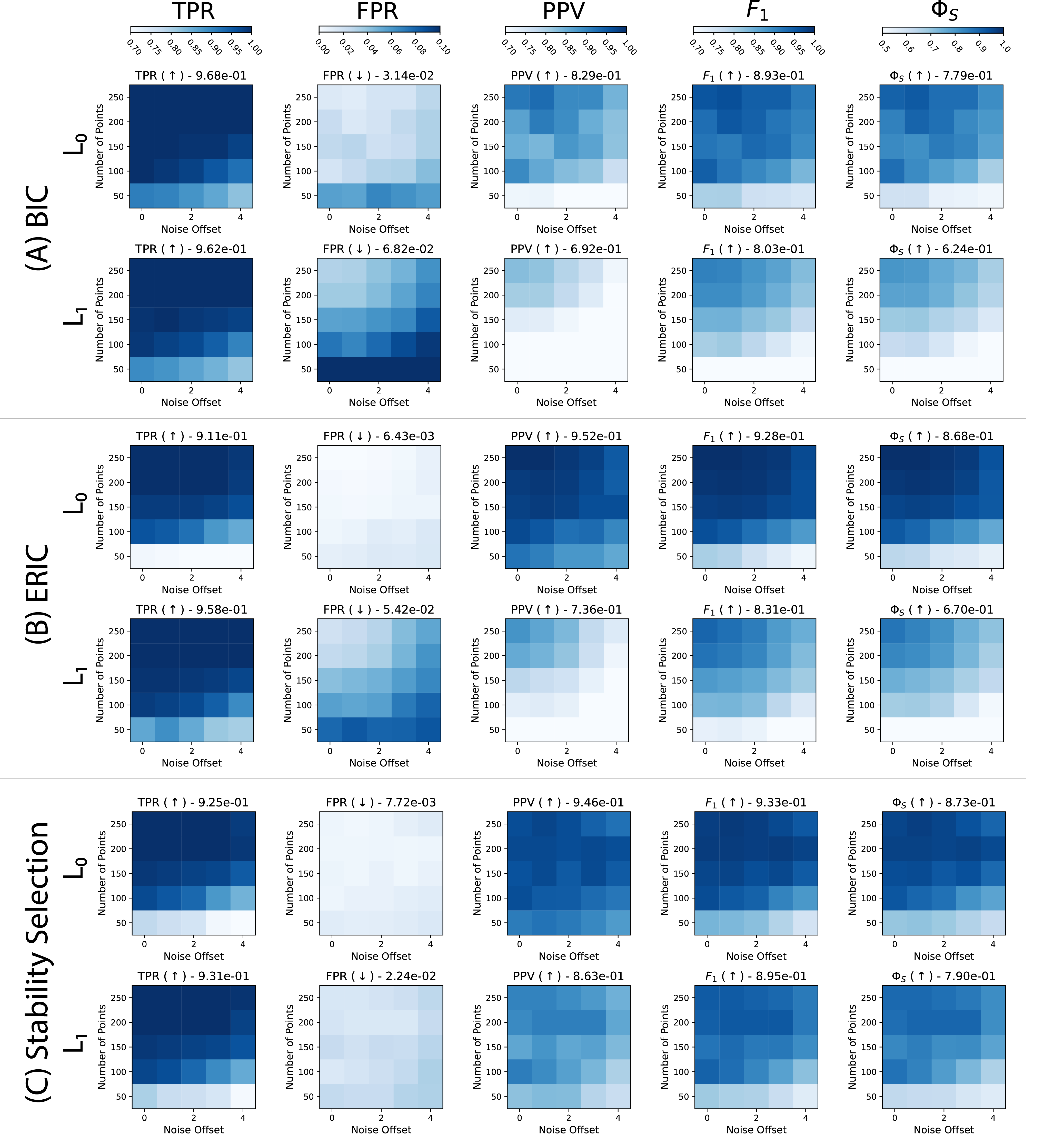}
  \caption{Variable-selection performance for a Poisson point process with localization uncertainty (scenario \textbf{P1}). We show the mean (over 100 independent repetitions of each experiment) True Positive Rate (TPR), False Positive Rate (FPR), Positive Predictive Value (PPV), $F_1$ score, and feature-selection stability $\Phi_S$ for model selection using the BIC \textbf{(A)}, ERIC \textbf{(B)}, and stability selection with $\mathrm{PFER}\leq 1$ \textbf{(C)} with adaptive $L_0$ (top row of each subfigure) and adaptive $L_1$ (bottom row of each subfigure) penalties. Each panel shows a performance metric (top titles, color bars) for different noise magnitudes $c$ ($x$-axis) and sample sizes $\mathbb{E}N(W)$ ($y$-axis). The average metrics over all 25 experiments are given in the panel titles with arrows ($\uparrow/\downarrow$) indicating the direction of improvement.}
  \label{fig:results:simulation:poisson:offset_metrics}
\end{figure}

Figure~\ref{fig:results:simulation:poisson:offset_metrics} shows the performance of the considered algorithms for scenario \textbf{P1} (Poisson point process with localization uncertainty). It compares model selection based on the BIC (A), ERIC (B), and stability selection with $\mathrm{PFER}\leq 1$ (C) for the adaptive $L_0$ (top row of each subfigure) and adaptive $L_1$ (bottom row of each subfigure) penalties. Throughout all cases, performance decreases for lower sample sizes and higher noise magnitudes, as sampling bias increases or the true covariates are masked. The adaptive $L_0$ penalty outperforms the adaptive $L_1$ penalty in all metrics except in the TPR, where the adaptive $L_1$ penalty achieves better results in particular for small sample sizes. This is because the adaptive $L_1$ penalty selects more covariates than the adaptive $L_0$ penalty. Choosing the penalty therefore allows tuning the tradeoff between TPR and FPR depending on the incurred costs of type I and type II errors, respectively. The $F_1$ score and the stability $\Phi_S$, however, are always better when using the adaptive $L_0$ penalty, despite its non-convex nature, as it reflects the true variable-selection objective.

Comparing selection strategies in Fig.~\ref{fig:results:simulation:poisson:offset_metrics}A-C, we note that the BIC achieves the lowest performance in all metrics except the TPR, for the same reasons as discussed above. As already reported by \citep{ba_inference_2023}, ERIC performs better than BIC, since it includes information about the regularization weight $\lambda$. ERIC achieves particularly good performance in combination with the adaptive $L_0$ penalty. For small sample sizes, however, this combination sometimes selects empty models (low TPR) because the likelihoods are attenuated by down-weighting the degrees of freedom for large $\lambda$. Stability selection achieves the best $F_1$ scores with $L_0$ performance comparable to the ERIC. This highlights that stronger penalization alone can already improve variable-selection performance. Also for stability selection, some empty models were selected for the smallest sample size and the highest noise level, where no covariate reached the threshold $\pi_{\mathrm{th}}=0.9$. However, this behavior is correct in light of the imposed error bound.

\begin{figure}
  \centering
  \includegraphics[width=0.5\linewidth]{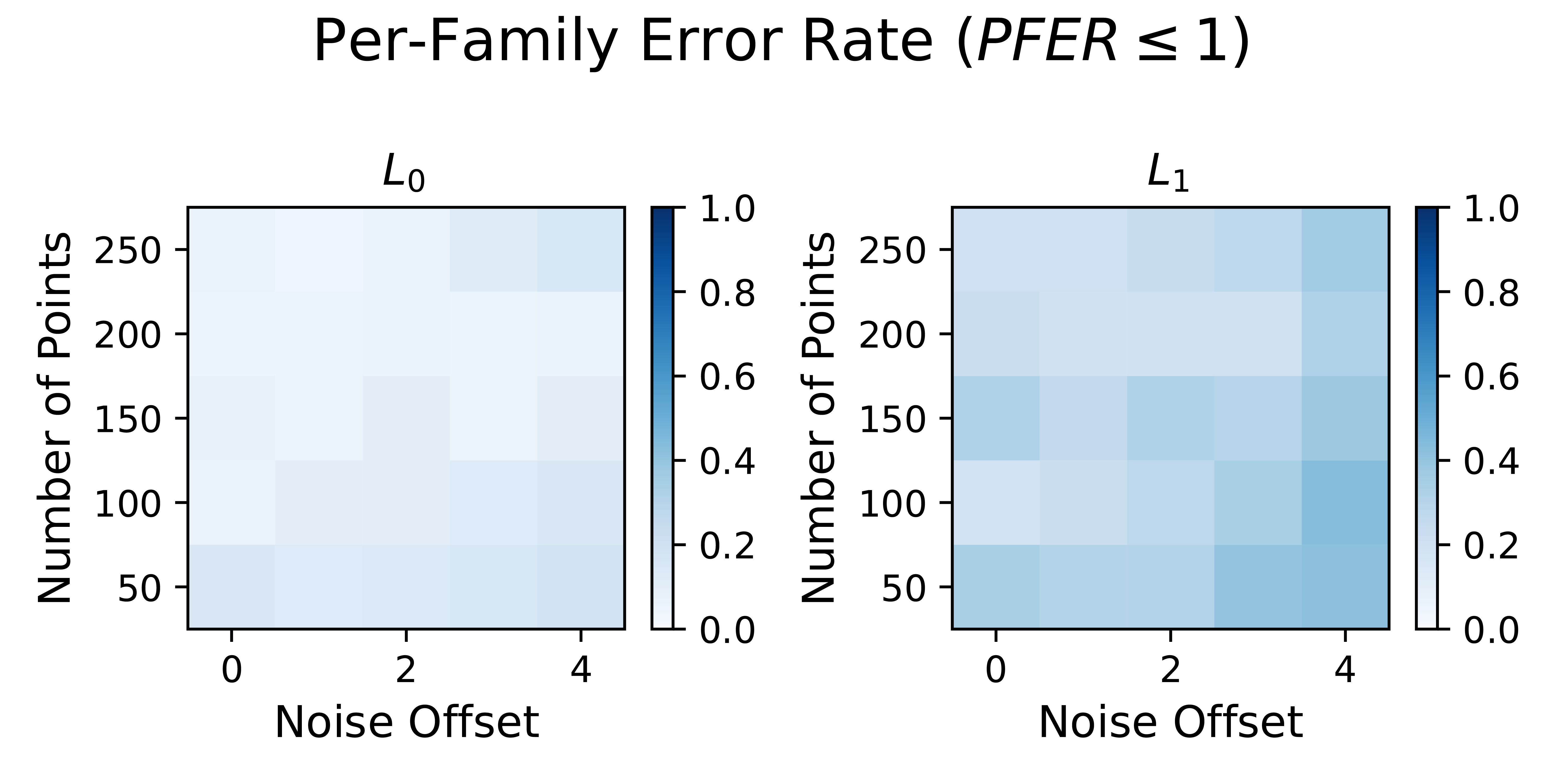}
  \caption{Empirical confirmation that stability selection achieves the desired error bound $\mathrm{PFER}\leq 1$ for all experiments in scenario \textbf{P1}.}
  \label{fig:results:simulation:poisson:pfer}
\end{figure}

The greatest difference between stability selection and ERIC is observed for the adaptive $L_1$ penalty. There, stability selection  reduces the FPR and increases the $F_1$ score. This is because the PFER bound leads to fewer selected covariates. In Fig.~\ref{fig:results:simulation:poisson:pfer} we show the empirically achieved PFER in all cases for both the adaptive $L_0$ and $L_1$ penalties. It is always well below the imposed bound of 1, confirming that stability selection is able to bound the number of false positives.

In order to derive scientific conclusions from the selected models, it is
key that variable selection be stable under noise, i.e., that the same covariates are reproducibly selected for different realizations of the process. We quantify this by the feature-selection stability $\Phi_S$ as defined in Eq.~\eqref{eq:methods:selection:stability_metric}. The results are shown in the last column of Fig.~\ref{fig:results:simulation:poisson:offset_metrics}. The adaptive $L_0$ penalty achieves better stability than the adaptive $L_1$ penalty. This is expected, as $L_0$ selects fewer covariates and thus has a lower variance in the selection indicator. Using stability selection instead of information criteria generally improves the stability for both penalties, especially at low sample sizes and high noise levels. For high sample sizes ERIC with $L_0$ penalization performs best, indicating that in this case strong penalization helps to stabilize the selection. The bootstrap procedure of stability selection implicitly improves the stability of the selected variables as previously reported \citep{nogueira_stability_2018}. This improved variable-selection stability is particularly important in the presence of noise.

\begin{figure}
  \centering
  \includegraphics[width=\linewidth,trim={0 0 0.5cm 0},clip]{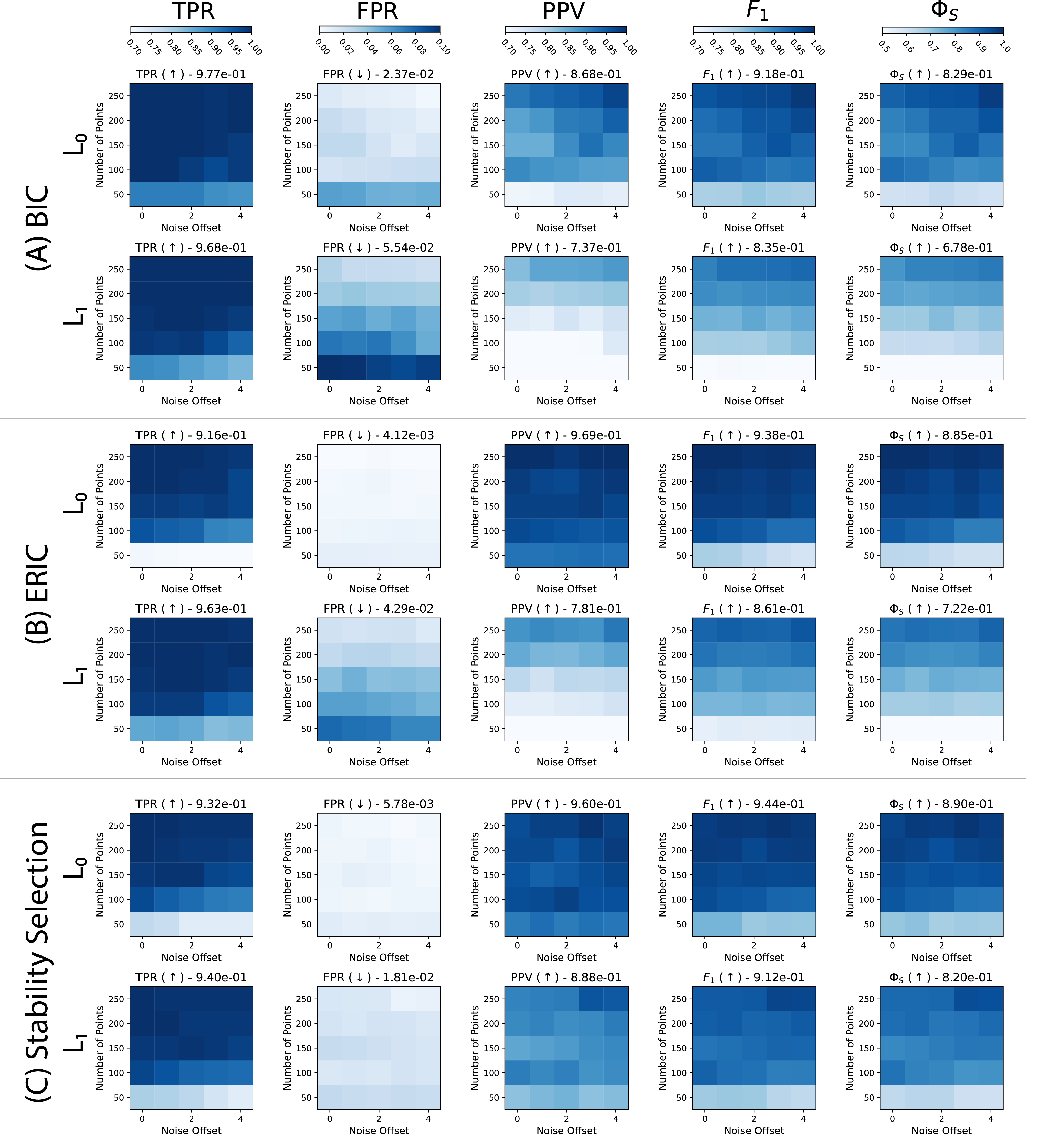}
  \caption{Variable-selection performance for a Poisson point process with detection uncertainty (scenario \textbf{P2}). We show the mean (over 100 independent repetitions of each experiment) True Positive Rate (TPR), False Positive Rate (FPR), Positive Predictive Value (PPV), $F_1$ score, and feature-selection stability $\Phi_S$ for model selection using the BIC \textbf{(A)}, ERIC \textbf{(B)}, and stability selection with $\mathrm{PFER}\leq 1$ \textbf{(C)} with adaptive $L_0$ (top row of each subfigure) and adaptive $L_1$ (bottom row of each subfigure) penalties. Each panel shows a performance metric (top titles, color bars) for different noise magnitudes $c$ ($x$-axis) and sample sizes $\mathbb{E}N(W)$ ($y$-axis). The average metrics over all 25 experiments are given in the panel titles with arrows ($\uparrow/\downarrow$) indicating the direction of improvement.}
  \label{fig:results:simulation:poisson:thinning_metrics}
\end{figure}

We next consider scenario \textbf{P2}, a Poisson point process with detection uncertainty. The results are reported in Fig.~\ref{fig:results:simulation:poisson:thinning_metrics} in the same format. They are qualitatively similar to those for scenario \textbf{P1} (Fig.~\ref{fig:results:simulation:poisson:offset_metrics}) but with better average performance. Again, the adaptive $L_0$ penalty outperforms the adaptive $L_1$ penalty in all metrics except the TPR, as it favors higher sparsity. Also for this noise type, stability selection improves the performance for both penalties, but particularly for the adaptive $L_1$ penalty. The PFER bound is met in all cases also here (not shown), and stability selection again has better $\Phi_S$ than the information criteria, in particular for high noise magnitudes. We therefore conclude that the adaptive $L_0$ penalty in combination with stability selection achieves the best variable-selection performance for uncorrelated Poisson point processes under noise, for both noise types.

In scenario \textbf{P1}, performance monotonically decreases with increasing noise magnitude. Curiously, in scenario \textbf{P2}, we sometimes observe better performance for higher noise magnitudes. While this seems counter-intuitive at first, it can be explained by thinning noise removing points from the process according to their local hardcore neighborhoods.
This introduces an apparent repulsive interaction between points, significantly influencing the second-order properties of the process \citep{kuronen_point_2021}. This is specific to the chosen noise model, and missing points by independent thinning would not have this effect \citep{moller_statistical_2003}.
The resulting regular process has fewer effective degrees of freedom (see Eq.~\eqref{eq:methods:model_selection:cbic:covariance} for $g(r)\leq 1$), allowing more stable estimation of the parameters. This is also reflected in the stability scores being higher for thinning noise (scenario \textbf{P2}, Fig.~\ref{fig:results:simulation:poisson:thinning_metrics}) than for displacement noise (scenario \textbf{P1}, Fig.~\ref{fig:results:simulation:poisson:offset_metrics}). This exemplifies that noise can influence variable selection in non-trivial ways that can be qualitatively different for different noise processes.

After having established the baseline for uncorrelated Poisson processes, we consider a Thomas point process with spatial attraction leading to clustering of points. We again consider both localization uncertainty (scenario \textbf{T1}) and detection uncertainty (scenario \textbf{T2}).

\begin{figure}
  \centering
  \includegraphics[width=\linewidth,trim={0 0 0.5cm 0},clip]{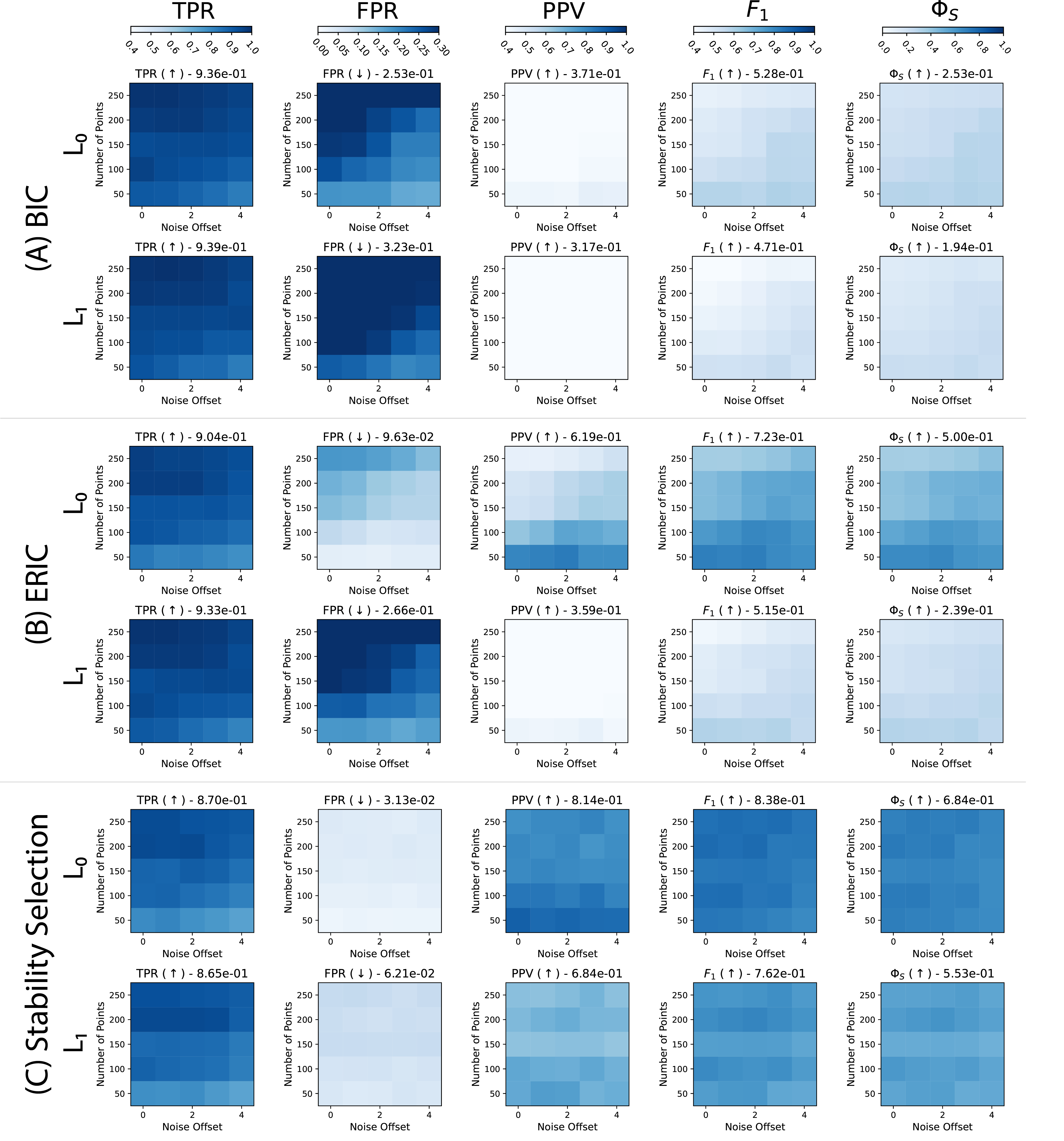}
  \caption{Variable-selection performance for a Thomas point process with localization uncertainty (scenario \textbf{T1}). We show the mean (over 100 independent repetitions of each experiment) True Positive Rate (TPR), False Positive Rate (FPR), Positive Predictive Value (PPV), $F_1$ score, and feature-selection stability $\Phi_S$ for model selection using the BIC \textbf{(A)}, ERIC \textbf{(B)}, and stability selection with $\mathrm{PFER}\leq 1$ \textbf{(C)} with adaptive $L_0$ (top row of each subfigure) and adaptive $L_1$ (bottom row of each subfigure) penalties. Each panel shows a performance metric (top titles, color bars) for different noise magnitudes $c$ ($x$-axis) and sample sizes $\mathbb{E}N(W)$ ($y$-axis). The average metrics over all 25 experiments are given in the panel titles with arrows ($\uparrow/\downarrow$) indicating the direction of improvement.}
  \label{fig:results:simulation:thomas:offset_metrics}
\end{figure}

The results are shown in Fig.~\ref{fig:results:simulation:thomas:offset_metrics} for scenario \textbf{T1}. The performance is generally lower than for a Poisson process, which is expected for the more complex Thomas process. Again, the adaptive $L_0$ penalty achieves a better performance than the adaptive $L_1$ penalty in all metrics except the TPR. Also as in the Poisson scenarios, the BIC achieves the overall lowest performance, followed by ERIC, which again works particularly well in conjunction with the adaptive $L_0$ penalty (37\% improvement in the average $F_1$ score over all experiments). Both BIC and ERIC achieve high TPR but low PPV, suggesting under-penalization of the clustering Thomas process. Stability selection achieves the best $F_1$ scores and feature-selection stability $\Phi_S$ for both $L_0$ and $L_1$ penalties. The PFER is again below the threshold in all cases (not shown), albeit much closer to 1 in the case of the $L_1$ penalty and even reaching the bound for $\mathbb{E}N(W)=250$ and $c=1$. This becomes apparent in the reduced FPR compared to BIC/ERIC, especially for the adaptive $L_1$ penalty. The adaptive $L_0$ penalty further improves the FPR over the adaptive Lasso. For both penalties, the FPR achieved by stability selection for the Thomas process is comparable to scenarios \textbf{P1} and \textbf{P2}. This suggests that the adaptive $L_0$ penalty in combination with stability selection is able to achieve high variable-selection performance without requiring knowledge about the second-order structure of the point process.

The performance of all methods tends to improve for smaller sample sizes or higher noise magnitudes, except for the TPR.
This counter-intuitive behavior is because the clustering in the Thomas process decreases with increasing noise, or it is not observed at low sample sizes. At high noise or for small samples, the Thomas process therefore appears more Poisson-like. The effect is particularly strong for the information criteria, which become increasingly appropriate the more Poisson-like the process appears. This provides another example where the effect of noise (this time simple displacement noise) on variable selection is not straightforward. Stability selection is the least sensitive to this effect, showing a more uniform distribution of performance metrics versus noise and sample size (Fig.~\ref{fig:results:simulation:thomas:offset_metrics}C). This is because stability selection does not model the noise process but generally reduces its influence.

Since the Thomas process introduces correlations between events, estimating its intensity
using the Poisson likelihood constitutes a composite likelihood estimate (see Section~\ref{sec:selection}).
We therefore also compare stability selection with the cBIC and cERIC, which have been shown to outperform the standard BIC and ERIC for clustering point processes \citep{choiruddin_information_2021}. We compute the effective degrees of freedom using Eq.~\eqref{eq:methods:model_selection:cbic:df} in two different ways: (1) Using the true pair-correlation function of the Thomas process with known parameters. While this is unrealistic in practical applications, it constitutes the best case for composite information criteria and hence serves us as a baseline. (2) By estimating the parameters of the $K$-function, the structure of which is analytically known, using minimum-contrast estimation. For this estimation, we use a two-step procedure \citep{waagepetersen_two-step_2009} with $r_{\min}=0$, $r_{\max}=25$, and $b=0.25$ to reduce the variance for clustering point processes as recommended by \citep{diggle_statistical_2013}. We use $K$-function estimation because it is less susceptible to noise than directly estimating the pair-correlation function. The latter would additionally require choosing a bandwidth parameter and would exhibit high variance. Moreover, we find that likelihood-based approaches, such as those proposed by \citep{waagepetersen_estimating_2007}, can become unstable in the presence of noise or for small sample sizes when clustering is weak. The minimum-contrast estimate is computed using the L-BFGS-B optimizer as implemented in the \texttt{scipy} Python package \citep{virtanen_scipy_2020}. The resulting parameter estimates are then used to compute the pair-correlation function required by composite information criteria with the variance--covariance matrix computed on a two-fold coarsened grid.

\begin{figure}
  \centering
  \includegraphics[width=\linewidth,trim={0 0 0.5cm 0},clip]{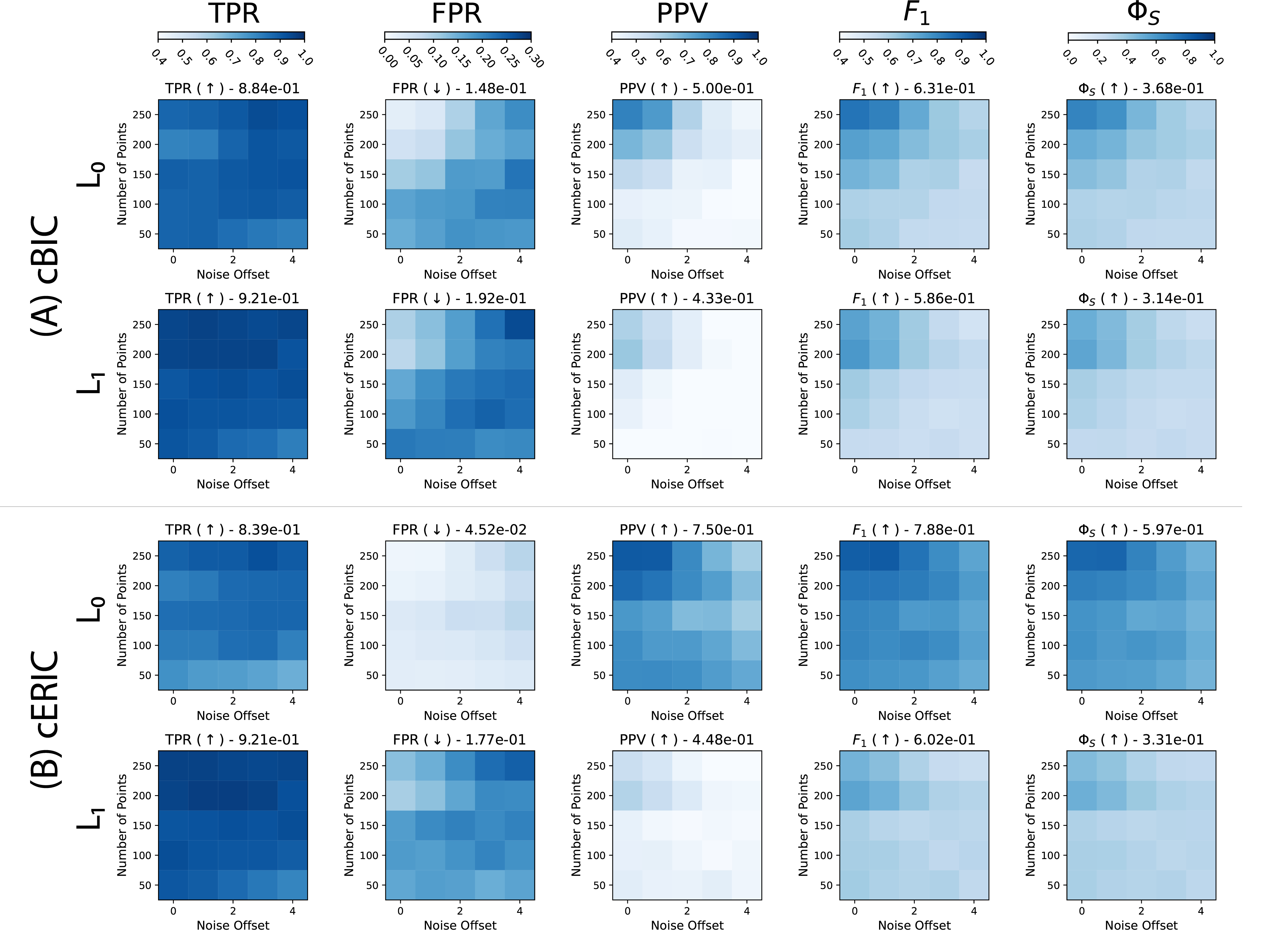}
  \caption{Variable-selection performance for a Thomas point process with localization uncertainty (scenario \textbf{T1}) using composite information criteria with parameter estimation. We show the mean (over 100 independent repetitions of each experiment) True Positive Rate (TPR), False Positive Rate (FPR), Positive Predictive Value (PPV), $F_1$ score, and feature-selection stability $\Phi_S$ for model selection using the cBIC \textbf{(A)} and cERIC \textbf{(B)} with adaptive $L_0$ (top row of each subfigure) and adaptive $L_1$ (bottom row of each subfigure) penalties. The parameters of the pair-correlation function are estimated using minimum-contrast estimation of the $K$-function. Each panel shows a performance metric (top titles, color bars) for different noise magnitudes $c$ ($x$-axis) and sample sizes $\mathbb{E}N(W)$ ($y$-axis). The average metrics over all 25 experiments are given in the panel titles with arrows ($\uparrow/\downarrow$) indicating the direction of improvement.}
  \label{fig:results:simulation:thomas:offset_metrics_cic_kest}
\end{figure}

\begin{figure}
  \centering
  \includegraphics[width=\linewidth,trim={0 0 0.5cm 0},clip]{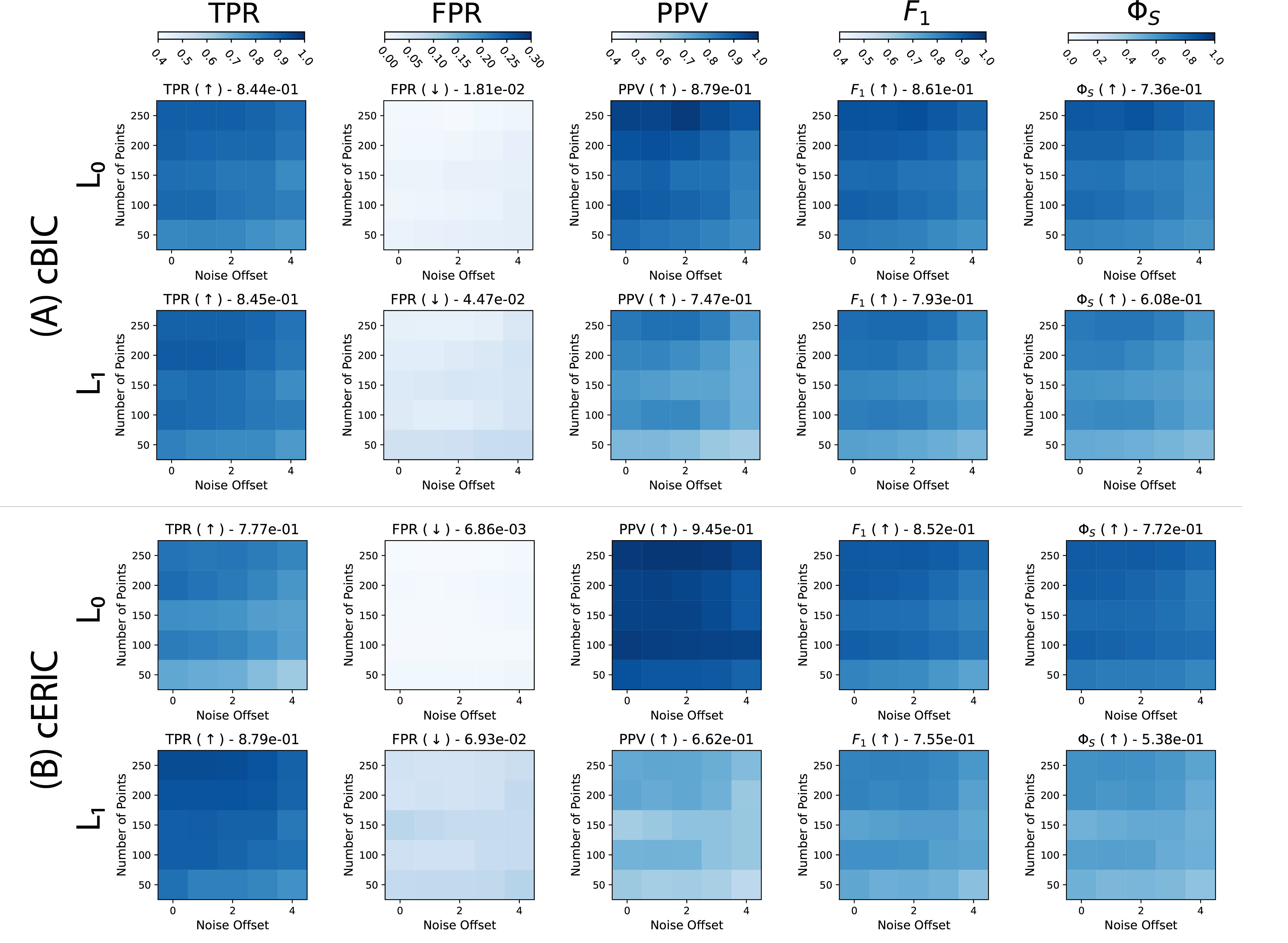}
  \caption{Variable-selection performance for a Thomas point process with localization uncertainty (scenario \textbf{T1}) using composite information criteria with exact knowledge. We show the mean (over 100 independent repetitions of each experiment) True Positive Rate (TPR), False Positive Rate (FPR), Positive Predictive Value (PPV), $F_1$ score, and feature-selection stability $\Phi_S$ for model selection using the cBIC \textbf{(A)} and cERIC \textbf{(B)} with adaptive $L_0$ (top row of each subfigure) and adaptive $L_1$ (bottom row of each subfigure) penalties. The parameters of the pair-correlation function are assumed to be known exactly. Each panel shows a performance metric (top titles, color bars) for different noise magnitudes $c$ ($x$-axis) and sample sizes $\mathbb{E}N(W)$ ($y$-axis). The average metrics over all 25 experiments are given in the panel titles with arrows ($\uparrow/\downarrow$) indicating the direction of improvement.}
  \label{fig:results:simulation:thomas:offset_metrics_cic}
\end{figure}

Figure~\ref{fig:results:simulation:thomas:offset_metrics_cic_kest} shows the results for cBIC and cERIC when using minimum-contrast estimation.
Both composite information criteria perform better than their uncorrected versions in all cases (Fig.~\ref{fig:results:simulation:thomas:offset_metrics}). Adaptive $L_0$ penalization achieves better performance and stability than the Lasso. However, neither cBIC nor cERIC achieve the performance of stability selection (Fig.~\ref{fig:results:simulation:thomas:offset_metrics}C), except for cERIC at low noise and large sample sizes. In these cases the estimation of the $K$-function is expected to work best and the stronger penalization leads to improved performance.
Overall, the average $F_1$ score across all experiments is 6\% higher for stability selection than for cERIC when using the $L_0$ penalty and 26\% higher for the $L_1$ penalty. For cBIC, the differences are even larger. We also observe that when using minimum-contrast estimation, composite information criteria increasingly select empty models for small sample sizes when using the $L_0$ penalty. We believe this is because the $K$-function overestimates the clustering due to the higher variance in the data and the renormalization by the estimated intensity function. In combination with the stronger $L_0$ penalty, this can lead to an over-penalization, favoring the homogeneous model where clustering is explained by the estimated $K$-function. This illustrates the potentially detrimental effect of feedback between the first- and second-order models in the estimation procedure, which can lead to suboptimal variable selection in the presence of noise. Such feedback is not present in stability selection, as it directly estimates the parameters of the intensity function directly from the data without requiring a second-order model.

Figure \ref{fig:results:simulation:thomas:offset_metrics_cic} shows the performance of cBIC and cERIC when the pair-correlation function is assumed to be known exactly. The performance is always better than when estimating the second-order parameters from data. This is expected, as the estimation procedure introduces additional errors, especially for small samples or high noise where the clustering is less apparent in the data. We also observe that with perfect knowledge, cBIC performs better than cERIC in the average $F_1$ score but achieves lower selection stability $\Phi_S$, indicating higher variance in the selected models. This is a consequence of the additional regularization in cERIC, through the penalization weight $\lambda$, leading to more conservative selection. Indeed, the higher PPV of cERIC indicates that it selects only the most relevant covariates, ignoring smaller effects. This could also explain the higher feature-selection stability. Overall, cBIC achieves a better trade-off in this setting, which explains its higher average $F_1$ score. When using the Lasso, PPV drops for cERIC. In this case, cERIC selects lower $\lambda$ values than cBIC, which leads to more complex models due to the soft thresholding of the Lasso. Similar observations have been made when using the adaptive Lasso and cERIC for Gibbs point processes \citep{ba_inference_2023}. However, this effect seems to be mitigated by the $L_0$ penalty, which leads to less smooth coefficient paths than the Lasso (see Fig.~\ref{fig:results:simulation:poisson:paths}). In the $\lambda$-ranges over which the estimated model does not change due to the hard thresholding of the $L_0$ penalty, cERIC can better prioritize models with higher penalization due to the $\lambda$-weighting.

Even in this best-case scenario, stability selection performs comparably with cBIC and cERIC.
The average $F_1$ score of stability selection is 3\% lower than for cBIC when using the $L_0$ penalty and around 4\% lower with the $L_1$ penalty. Feature-selection stability is between 8\% and 12\% lower than for cBIC and cERIC. For the $L_1$ penalty, stability selection performs better than cERIC (by 1\%) but worse than cBIC (by 4\%).
This, in combination with the higher TPR and smaller PPV of stability selection, suggests that stability selection produces more false positives than composite information criteria with perfect knowledge of the second-order structure of the process. In practical applications, where perfect second-order knowledge is not available, the performance of composite information criteria rapidly deteriorates. Even for high-quality estimates from data (Fig.~\ref{fig:results:simulation:thomas:offset_metrics_cic_kest}, parameter estimate for ground-truth analytical form), cBIC and cERIC fall behind. Stability selection therefore presents a good choice in practice. It remains robust to noise without requiring additional knowledge of the true process.

\begin{figure}
  \centering
  \includegraphics[width=\linewidth,trim={0 0 0.5cm 0},clip]{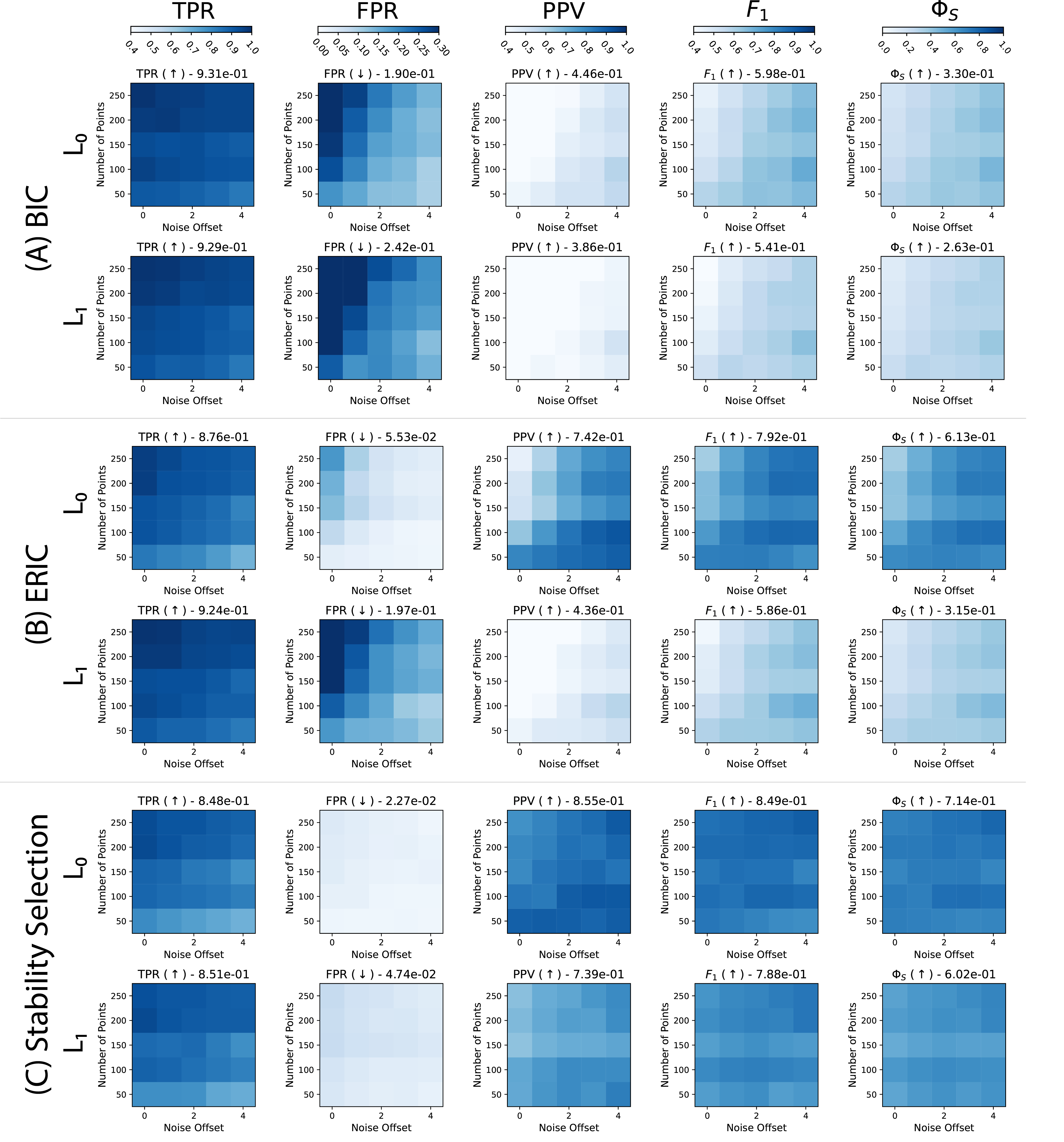}
  \caption{Variable-selection performance for a Thomas point process with detection uncertainty (scenario \textbf{T2}). We show the mean (over 100 independent repetitions of each experiment) True Positive Rate (TPR), False Positive Rate (FPR), Positive Predictive Value (PPV), $F_1$ score, and feature-selection stability $\Phi_S$ for model selection using BIC \textbf{(A)}, ERIC \textbf{(B)}, and stability selection with $\mathrm{PFER}\leq 1$ \textbf{(C)} with adaptive $L_0$ (top row of each subfigure) and adaptive $L_1$ (bottom row of each subfigure) penalties. Each panel shows a performance metric (top titles, color bars) for different noise magnitudes $c$ ($x$-axis) and sample sizes $\mathbb{E}N(W)$ ($y$-axis). The average metrics over all 25 experiments are given in the panel titles with arrows ($\uparrow/\downarrow$) indicating the direction of improvement.}
  \label{fig:results:simulation:thomas:thinning_metrics}
\end{figure}

We repeat the same experiments for a Thomas point process with detection uncertainty (scenario \textbf{T2}). The results are shown in Fig.~\ref{fig:results:simulation:thomas:thinning_metrics}. Like for scenario \textbf{T1}, the overall performance is again lower than in the Poisson case. In contrast to \textbf{T1}, but similar to \textbf{P2}, variable-selection performance is better than for localization uncertainty, and it generally improves with increasing detection noise magnitude. We hypothesize that this is again because thinning creates a more regular process. This again particularly impacts BIC and ERIC, with ERIC achieving better performance than BIC. The best performance overall is achieved by stability selection with $L_0$ penalty. Stability selection is also the least sensitive to noise and sample size.
Across methods, the adaptive $L_0$ penalty performs better in all metrics than the adaptive $L_1$ penalty, except for the TPR. This is because the adaptive Lasso includes more covariates overall, also leading to higher FPRs. Using stability selection instead of information criteria, however, reduces the FPR of the Lasso by an order of magnitude while also reducing the variance in the estimated models.

\begin{figure}
  \centering
  \includegraphics[width=\linewidth,trim={0 0 0.5cm 0},clip]{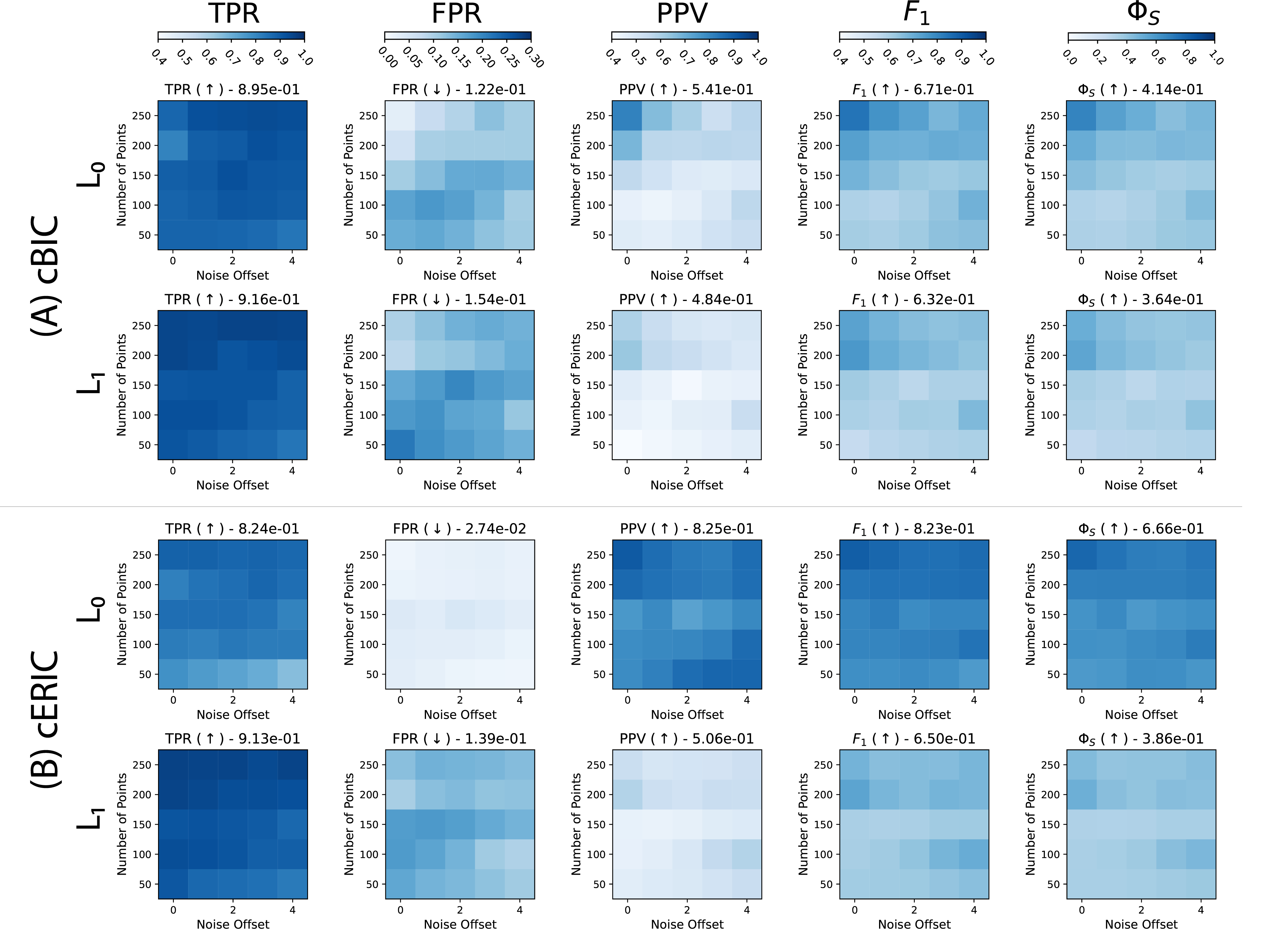}
  \caption{Variable-selection performance for a Thomas point process with detection uncertainty (scenario \textbf{T2}) using composite information criteria with parameter estimation. We show the mean (over 100 independent repetitions of each experiment) True Positive Rate (TPR), False Positive Rate (FPR), Positive Predictive Value (PPV), $F_1$ score, and feature-selection stability $\Phi_S$ for model selection using the cBIC \textbf{(A)} and cERIC \textbf{(B)} with adaptive $L_0$ (top row of each subfigure) and adaptive $L_1$ (bottom row of each subfigure) penalties. The parameters of the pair-correlation function are estimated using minimum-contrast estimation of the $K$-function. Each panel shows a performance metric (top titles, color bars) for different noise magnitudes $c$ ($x$-axis) and sample sizes $\mathbb{E}N(W)$ ($y$-axis). The average metrics over all 25 experiments are given in the panel titles with arrows ($\uparrow/\downarrow$) indicating the direction of improvement.}
  \label{fig:results:simulation:thomas:thinning_metrics_cic_kest}
\end{figure}

\begin{figure}
  \centering
  \includegraphics[width=\linewidth,trim={0 0 0.5cm 0},clip]{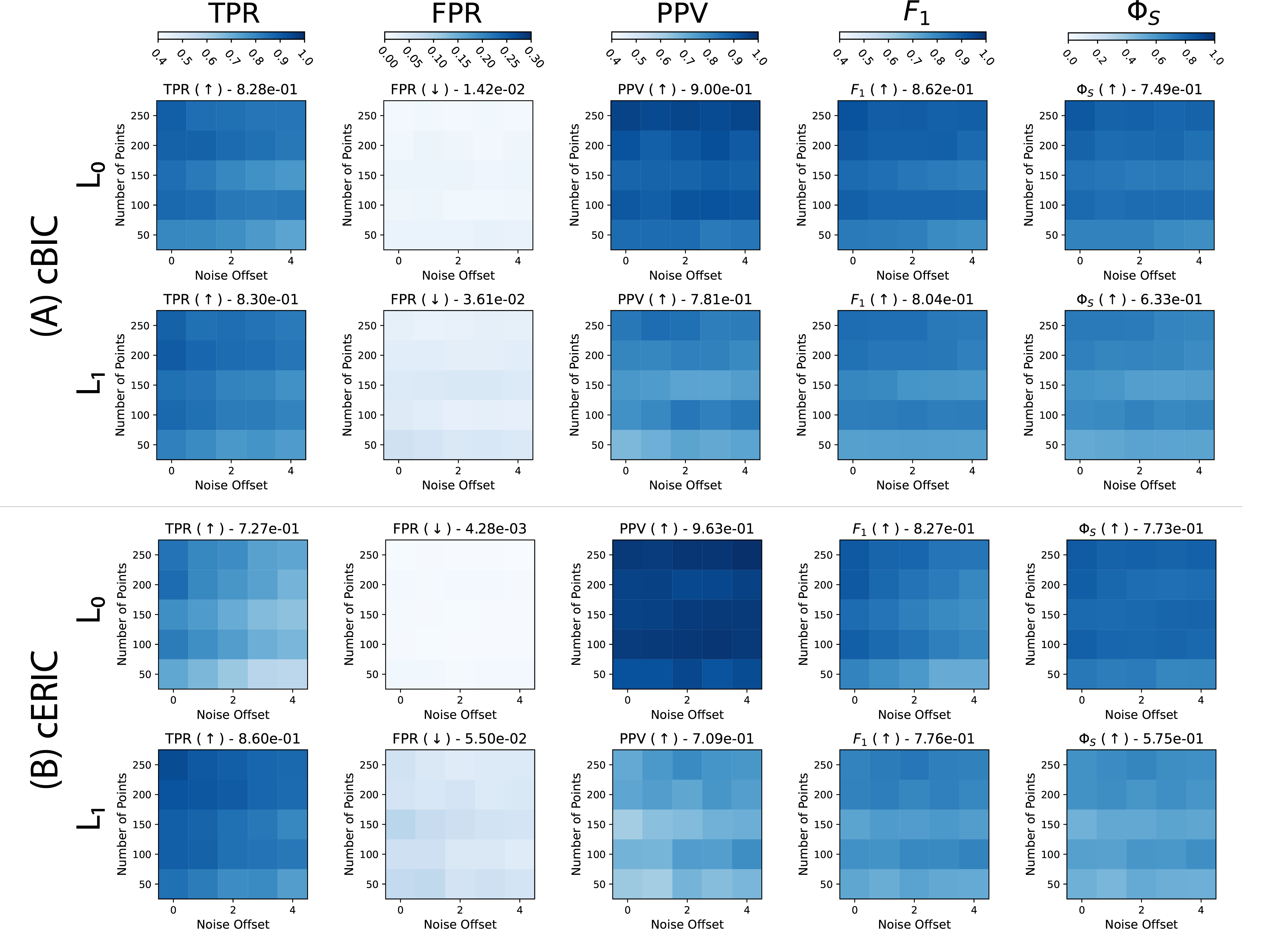}
  \caption{Variable-selection performance for a Thomas point process with detection uncertainty (scenario \textbf{T2}) using composite information criteria with exact knowledge. We show the mean (over 100 independent repetitions of each experiment) True Positive Rate (TPR), False Positive Rate (FPR), Positive Predictive Value (PPV), $F_1$ score, and feature-selection stability $\Phi_S$ for model selection using the cBIC \textbf{(A)} and cERIC \textbf{(B)} with adaptive $L_0$ (top row of each subfigure) and adaptive $L_1$ (bottom row of each subfigure) penalties. The parameters of the pair-correlation function are assumed to be known exactly. Each panel shows a performance metric (top titles, color bars) for different noise magnitudes $c$ ($x$-axis) and sample sizes $\mathbb{E}N(W)$ ($y$-axis). The average metrics over all 25 experiments are given in the panel titles with arrows ($\uparrow/\downarrow$) indicating the direction of improvement.}
  \label{fig:results:simulation:thomas:thinning_metrics_cic}
\end{figure}

Figure \ref{fig:results:simulation:thomas:thinning_metrics_cic_kest} shows the performance of cBIC and cERIC in scenario \textbf{T2} when the second-order parameters are estimated using minimum-contrast estimation. While thinning noise increases the local regularity of the point pattern, which tends to underestimate pair correlations, we still observe an improvement in performance when using composite information criteria over standard information criteria. This is especially visible for the cBIC, where the average $F_1$ score is around 12\% or 17\% higher than for BIC with $L_0$ or $L_1$ penalties, respectively. The cERIC achieves an improvement over the ERIC of 4\% and 11\%, respectively, for the $L_0$ and $L_1$ penalties. However, no composite information criterion achieves the performance or robustness of stability selection. While the $F_1$ score of cERIC with $L_0$ penalty is comparable to that of stability selection, stability selection achieves higher feature-selection stability $\Phi_S$.

Like for the previous scenario \textbf{T1}, we repeat the analysis with perfect knowledge of the pair-correlation function of the Thomas process under detection uncertainty. This constitutes the best case for cBIC and cERIC.
The results are shown in Fig.~\ref{fig:results:simulation:thomas:thinning_metrics_cic}. Again, like in scenario \textbf{T1}, performance is better than when using a data-estimated pair-correlation function. The cBIC again outperforms cERIC in terms of the average $F_1$ score for both penalties. Like for localization uncertainty, the $\lambda$-weighting in cERIC leads to a more conservative selection of covariates under the $L_0$ penalty and less penalization under the $L_1$ penalty. Also in this case, stability selection achieves near-best performance without requiring knowledge of the pair correlation function.

\subsection{Application to a forestry data set}

To illustrate the practical applicability of the proposed method, we consider the well-known data from the Barro Colorado Island (BCI) research plot in Panama. Over a 50\,ha site (1000\,m $\times$ 500\,m), the locations of tree stems with at least 1\,cm diameter at breast height have been recorded. This results in a data set of over 350,000 trees from 300 species \citep{condit_tropical_1998,hubbell_light-gap_1999}. A central question is how so many species are able to coexist, and how they carve environmental niches. Therefore, one aims to identify environmental covariates---such as elevation or soil nutrients---that explain the distribution of tree species. This data set has already been used in previous studies to identify sparse sets of predictors \citep{choiruddin_convex_2018,ba_inference_2023}, which allows direct comparison of results.

\begin{table}[]
  \caption{Stability-selection results for the BCI data set using adaptive Lasso ($L_1$) and adaptive Best-Subset ($L_0$) penalties. The table shows the estimated effect sizes (standardized data) for $\mathrm{PFER}\leq 1,2,3$. The effect sizes are estimated over the support identified by stability selection by solving the unpenalized composite likelihood problem on the whole data set. The considered covariates are: elevation (Elev.), slope (Slope), aluminum (Al), boron (B), calcium (Ca), copper (Cu), iron (Fe), potassium (K), magnesium (Mg), manganese (Mn), phosphorus (P), zinc (Zn), nitrogen (N), mineralized nitrogen (N(min)), and soil pH (pH). The last row shows the number of selected covariates $\big\Vert\hat\beta\big\Vert_0=\sum_{j=1}^p \mathbf{1}(\hat\beta_j \neq 0)$ for the respective penalty and error bound.}
  \centering\footnotesize
  \begin{tabular}{lcccccc}
    \toprule
    & \multicolumn{3}{c}{Adaptive Lasso ($L_1$)} & \multicolumn{3}{c}{Adaptive Best Subset ($L_0$)}  \\
    \cmidrule(lr){2-4} \cmidrule(lr){5-7}
    & $\mathrm{PFER}\leq 1$ & $\mathrm{PFER}\leq 2$ & $\mathrm{PFER}\leq 3$ & $\mathrm{PFER}\leq 1$ & $\mathrm{PFER}\leq 2$ & $\mathrm{PFER}\leq 3$ \\
    \midrule
    Elev.    &  0.35  &  0.39  &  0.38  &  0.35  &  0.35  &  0.36  \\
    Slope    &  0.33  &  0.27  &  0.30  &  0.33  &  0.33  &  0.33  \\
    Al       &  0     &  0     &  0     &   0    &   0    &   0    \\
    B        &  0     &  0.40  &  0.22  &   0    &   0    &   0    \\
    Ca       &  0     &  0     &  0     &   0    &   0    &   0    \\
    Cu       &  0     &  0     &  0     &   0    &   0    &   0    \\
    Fe       &  0     &  0     &  0     &   0    &   0    &   0    \\
    K        &  0     &  0     &  0     &   0    &   0    &   0    \\
    Mg       &  0     &  0     &  0     &   0    &   0    &   0    \\
    Mn       &  0     &  0     &  0.26  &   0    &   0    &   0.33 \\
    P        &  -0.59 &  -0.68 &  -0.61 &  -0.59 &  -0.59 &  -0.54 \\
    Zn       &  -0.28 &  -0.58 &  -0.57 &  -0.28 &  -0.28 &  -0.45 \\
    N        &  0     &  0     &  0     &   0    &   0    &   0    \\
    N(min)   &  0     &  0     &  0     &   0    &   0    &   0    \\
    pH       &  0     &  0     &  0     &   0    &   0    &   0    \\
    \midrule
    $\big\Vert\hat\beta\big\Vert_0$ &  4   &  5  &  6  &  4  &  4  &  5 \\
    \bottomrule
  \end{tabular}
  \label{tab:results:forestry:variable_selection}
\end{table}

\begin{figure}
  \centering
  \includegraphics[width=\linewidth]{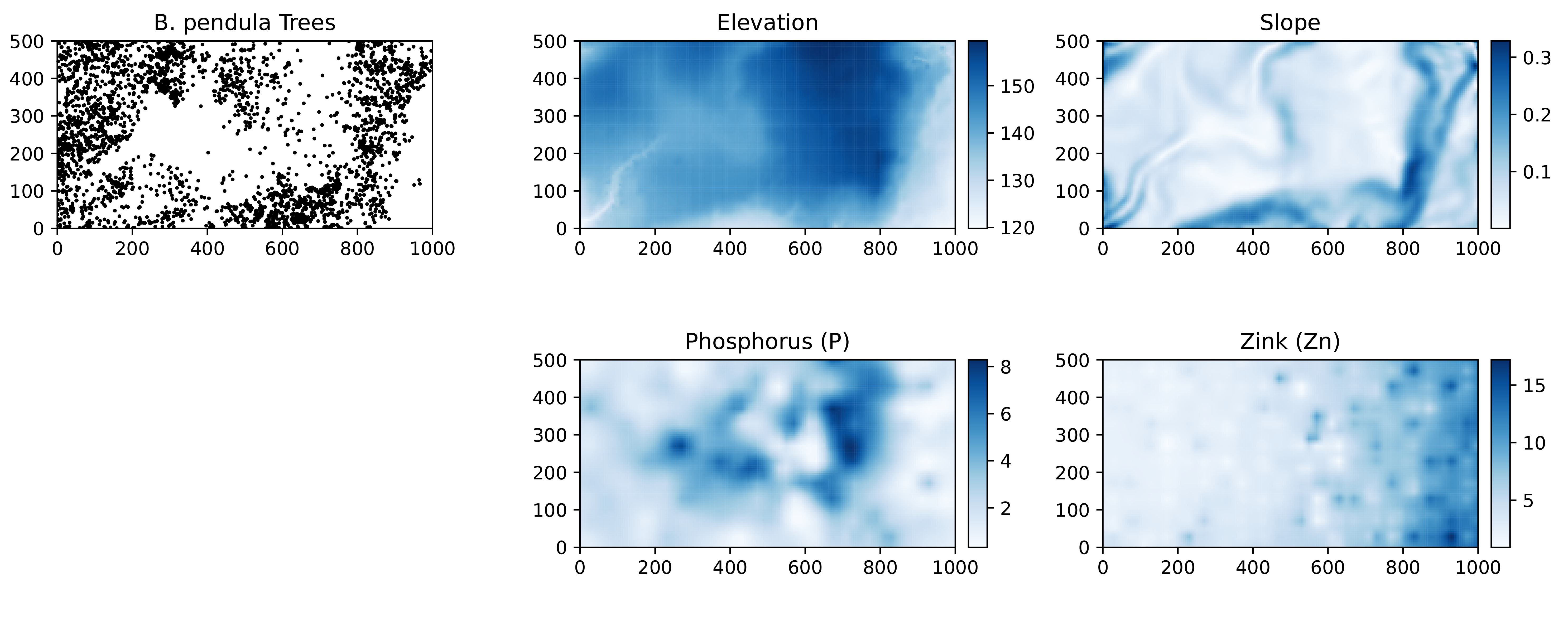}
  \caption{Point pattern of \textit{B.~pendula} tree stem locations in the BCI research plot (black dots, top-left) with visualizations of the selected covariates from Table~\ref{tab:results:forestry:variable_selection} for $\mathrm{PFER}\leq 1$. The covariates are: elevation (Elev.), slope (Slope), phosphorus (P), and zinc (Zn).}
  \label{fig:results:forestry:point_pattern}
\end{figure}

We focus on the locations of the 3,604 \emph{Beilschmiedia pendula} (BPL) trees from the \textit{Lauraceae} family, which we interpret as a realization of a spatial point process (see Fig.~\ref{fig:results:forestry:point_pattern}, top-left). We aim to model the intensity function of the point process using the 15 covariates listed in Table~\ref{tab:results:forestry:variable_selection}. As in Section~\ref{sec:results:simulation_study}, we interpolate all covariates onto a common grid of size $201\times101$, which we also use a quadrature points. The observation window corresponds to the entire 50\,ha region, $W=[0,1000]\times[0,500]$. All covariates are standardized as in previous analyses \citep{choiruddin_convex_2018,ba_inference_2023}.

We compare the models identified by stability selection with adaptive $L_1$ and adaptive $L_0$ penalties. The PGD step size is $\gamma=10^{-4}$ in both cases, which ensures numerical stability. For stability selection, we obtain 50 bootstrap samples by $p$-thinning with $p_{\mathrm{thin}}=0.5$. The $\lambda$-path is chosen so as to achieve a desired PFER bound according to Eq.~\eqref{eq:methods:selection:stability:error_control} with $\pi_{\mathrm{th}}=0.9$. The resulting candidate interval for $\lambda$ spans six orders of magnitude and is discretized with 40 log-equidistant points. The $\lambda_{\max}$ is set to obtain an empty model under the respective penalty. Using these settings, the stability path for the $L_1$ penalty is computed in 20s and for the $L_0$ penalty in 38s on a personal laptop (Apple MacBook Pro 2023, Apple M3 Pro CPU, 36\,GB LPDDR5 RAM). The higher computational cost of the $L_0$ penalty is due to the fixed step sizes in contrast to the BB-PGD used for the $L_1$ penalty. Table~\ref{tab:results:forestry:variable_selection} reports the resulting models for $\mathrm{PFER}\leq 1,2,3$ and compares the two penalization methods. The final parameter estimates are obtained by solving the unpenalized composite likelihood problem for the identified set of stable predictors using all observations.

With either penalty, stability selection yields similar models. As expected, the adaptive Lasso ($L_1$) selects slightly more covariates than adaptive Best-Subset selection ($L_0$). For $\mathrm{PFER}\leq 1$, both penalties find the same
covariates elevation (Elev.), slope (Slope), phosphorus (P), and zinc (Zn). This agrees with previous results on this data set \citep{choiruddin_convex_2018}. These covariates are visualized in Fig.~\ref{fig:results:forestry:point_pattern} together with the tree location pattern, visually supporting the conclusion that BPL trees prefer higher elevation and slopes with low concentration of phosphorus and zinc.

The influence of error control can nicely be seen in the number of selected covariates monotonically increasing for looser bounds and does not require recomputing the path. At higher $\mathrm{PFER}$, the $L_0$ models only slowly include additional terms, with the first one being manganese (Mn) at $\mathrm{PFER}\leq 3$. The $L_1$ models first and additionally include boron (B). In this case, the effect sizes are similar as previously reported using weighted composite likelihood estimates, which rely on estimating the pair-correlation function \citep{choiruddin_convex_2018}. Stability selection achieves the same level of sparsity without modeling assumptions about the underlying process, and without the need for estimating the pair-correlation function. The similarity of the obtained models hints at the practical utility of stability selection for variable selection in spatial point processes.

\section{Discussion and Outlook}\label{sec:discussion}

We presented a method for sparse intensity estimation of spatial point processes under noise. We showed that noise, in the form of localization or detection uncertainty, significantly influences variable-selection performance. We proposed a combination of adaptive best-subset selection ($L_0$ penalty) and stability selection to improve noise robustness without requiring additional knowledge about the true process. Simulation benchmarks showed that the proposed method provides more stable estimates and allows for error control in the variable-selection procedure. Together, this improves variable-selection performance under noise, for small samples, and under model misspecification.

The presented method uses $p$-thinning for generating the bootstrap subsamples required for stability selection. The estimating equation in Eq.~\eqref{eq:methods:selection:stability:estimating_equations} allows for straightforward integration of the presented method into existing frameworks for spatial point processes that use generalized linear model (GLM) software \citep{baddeley_spatial_2016}. The computational overhead incurred by our method scales linearly with the number of bootstrap samples $K$, where $K=50$ was sufficient in our benchmarks. While the compute time depends on many factors, including the number of covariates, the discretization grid size, the step size, the convergence criteria, and the quality of the warm starts when computing the $\lambda$-path, the times for the forestry dataset were below one minute per case. 

We compared the proposed method to existing model-selection procedures based on (composite) information criteria. The present method consistently outperformed information criteria both in terms of stability and accuracy. Typically, model selection based on information criteria produced more false positives.
For the correlated Thomas point process, stability selection performed almost as well as composite information criteria with perfect knowledge of the pair-correlation function of the process, albeit without requiring such knowledge. This suggests that stability selection is able to cope with simultaneous clustering and noise in the data without having to explicitly model pair correlations or noise statistics. This is of practical importance, since such knowledge is not usually available in applications, and it is difficult to estimate for small samples. But even when second-order information could be estimated from the data, the performance of composite information criteria markedly dropped. However, it was still better than using the standard BIC or ERIC, which we deem not advisable for correlated processes or noisy observations.

Our results also suggested that the adaptive $L_0$ penalty, despite its non-convexity, generally achieves better performance than the adaptive $L_1$ penalty at comparable computational cost, especially in terms of the false-positive rate and $F_1$ score. This indicates that proximal gradient descent is able to effectively find sparse local minima that represent good-enough estimates of the true process.
Choosing between $L_0$ and $L_1$ penalties allows trading off between type I and type II errors, since $L_0$ generally selects fewer variables. Stability selection provides control over the PFER for either penalization. The empirical PFER never exceeded the imposed threshold in all presented simulations.
This error control filters out bad local minima, leading to greatly reduced false-positive rates, particularly when using the adaptive Lasso. Existing estimators based on the Lasso can therefore directly benefit from stability selection.

Finally, we illustrated the applicability of the proposed method on a real-world forestry data set, identifying the relevant covariates for the spatial distribution of \emph{Beilschmiedia pendula} trees in a tropical rain forest area of 50\,ha. The identified models were in line with previous results, validating the method. Comparing different error bounds and penalties also confirmed the statistical consistency of the proposed method, with looser error bounds leading to monotonically larger models. This indicates that stability selection can be used to identify a sparse set of relevant covariates in spatial point processes without any assumptions on the underlying model.

While these results are encouraging, they also highlight the need for future work.
An obvious limitation of our work is that we only considered uncorrelated noise. Real-world data, however, often contains structured noise.
Examples include presence-only analysis of plants and animals in ecological studies, where they are more likely to be spotted near roads, or digital imagery recorded with CMOS or CCD sensors, which generate correlated readout noise.
The effect of structured noise on stability selection remains to be studied.

In order to establish a baseline, we here only considered the classic formulation of stability selection.
It would be interesting in the future to extend the presented approach to more recent stability-selection variants, such as complimentary-pairs stability selections \citep{shah_variable_2013}. It would also be worthwhile to explore the use of stability selection for estimating interactions across multiple spatial scales in multivariate point process models, particularly when combined with group sparsity as suggested by \citep{rajala_detecting_2018}. Group-sparse multivariate models are relevant in biology and ecology, where many (molecular) species interact with each other and with their environment across scales. Stability selection may offer a systematic approach to identifying interaction structures in such settings, while being robust to noise in the data.

Finally, future work could extend the methodology proposed here to other types of point processes. Since similar discretizations of composite likelihoods are, for example, also used for Gibbs point processes \citep{baddeley_practical_2000,ba_inference_2023}. We therefore think that the presented method could be adapted to estimating the Papangelou conditional intensity in those models as well.

Despite these open questions, we believe that the idea of applying stability selection to spatial point-process modeling opens several doors. We hope that the present work laid the foundations by systematically benchmarking the method, deriving the estimating equations, and establishing the effectiveness of proximal gradient descent for non-convex $L_0$ penalties.

\section*{Acknowledgments}
The authors thank Dr.~Nandu Gopan and Dr.~Abhishek Behera (both Sbalzarini group) for helpful discussions. The authors acknowledge financial support by the German Federal Ministry of Research, Technology and Space and by the Saxon State Ministry of Science, Culture, and Tourism in the program ``Center of Excellence for AI-research'' --- ``Center for Scalable Data Analytics and Artificial Intelligence Dresden/Leipzig'', project identification number: ScaDS.AI. This work was supported by the German Research Foundation (Deutsche Forschungsgemeinschaft, DFG) under Germany's Excellence Strategy --- Cluster of Excellence EXC2068  ``Physics of Life'' of TU Dresden.

\section*{Data Availability}
The data, models, and source codes that support the findings of this study are openly
available at the URL/DOI: \url{https://git.mpi-cbg.de/mosaic/point_process_stability_selection}.

\bibliographystyle{unsrt}

\begin{thebibliography}{10}

\bibitem{renner_equivalence_2013}
Ian~W. Renner and David~I. Warton.
\newblock Equivalence of {MAXENT} and {Poisson} {Point} {Process} {Models} for {Species} {Distribution} {Modeling} in {Ecology}.
\newblock {\em Biometrics}, 69(1):274--281, March 2013.

\bibitem{renner_point_2015}
Ian~W. Renner, Jane Elith, Adrian Baddeley, William Fithian, Trevor Hastie, Steven~J. Phillips, Gordana Popovic, and David~I. Warton.
\newblock Point process models for presence‐only analysis.
\newblock {\em Methods in Ecology and Evolution}, 6(4):366--379, April 2015.

\bibitem{waagepetersen_two-step_2009}
Rasmus Waagepetersen and Yongtao Guan.
\newblock Two-{Step} {Estimation} for {Inhomogeneous} {Spatial} {Point} {Processes}.
\newblock {\em Journal of the Royal Statistical Society Series B: Statistical Methodology}, 71(3):685--702, June 2009.

\bibitem{zimmerman_estimating_2008}
Dale~L. Zimmerman.
\newblock Estimating the {Intensity} of a {Spatial} {Point} {Process} from {Locations} {Coarsened} by {Incomplete} {Geocoding}.
\newblock {\em Biometrics}, 64(1):262--270, March 2008.

\bibitem{diggle_statistical_2013}
Peter~J. Diggle.
\newblock {\em Statistical {Analysis} of {Spatial} and {Spatio}-{Temporal} {Point} {Patterns}}.
\newblock Monographs on {Statistics} and {Applied} {Probability} 128. Chapman and Hall/CRC, New York, 3 edition, July 2013.

\bibitem{helmuth_beyond_2010}
Jo~A Helmuth, Grégory Paul, and Ivo~F Sbalzarini.
\newblock Beyond co-localization: inferring spatial interactions between sub-cellular structures from microscopy images.
\newblock {\em BMC Bioinformatics}, 11(1):372, December 2010.

\bibitem{parra_methods_2021}
Edwin~Roger Parra.
\newblock Methods to {Determine} and {Analyze} the {Cellular} {Spatial} {Distribution} {Extracted} {From} {Multiplex} {Immunofluorescence} {Data} to {Understand} the {Tumor} {Microenvironment}.
\newblock {\em Frontiers in Molecular Biosciences}, 8:668340, June 2021.

\bibitem{summers_spatial_2022}
Huw~D. Summers, John~W. Wills, and Paul Rees.
\newblock Spatial statistics is a comprehensive tool for quantifying cell neighbor relationships and biological processes via tissue image analysis.
\newblock {\em Cell Reports Methods}, 2(11):100348, November 2022.

\bibitem{li_statistical_2015}
Yingzhe Li, Francois Baccelli, Harpreet~S. Dhillon, and Jeffrey~G. Andrews.
\newblock Statistical {Modeling} and {Probabilistic} {Analysis} of {Cellular} {Networks} {With} {Determinantal} {Point} {Processes}.
\newblock {\em IEEE Transactions on Communications}, 63(9):3405--3422, September 2015.

\bibitem{hastie_statistical_2015}
Trevor Hastie, Robert Tibshirani, and Martin Wainwright.
\newblock {\em Statistical learning with sparsity: the lasso and generalizations}.
\newblock Number 143 in Monographs on statistics and applied probability. CRC Press, Taylor \& Francis Group, Boca Raton, 2015.

\bibitem{thurman_variable_2014}
Andrew~L. Thurman and Jun Zhu.
\newblock Variable selection for spatial {Poisson} point processes via a regularization method.
\newblock {\em Statistical Methodology}, 17:113--125, March 2014.

\bibitem{thurman_regularized_2014}
Andrew Thurman, Rao Fu, Yongtao Guan, and Jun Zhu.
\newblock Regularized {Estimating} {Equations} for {Model} {Selection} of {Clustered} {Spatial} {Point} {Processes}.
\newblock {\em Statistica Sinica}, 2014.

\bibitem{choiruddin_convex_2018}
Achmad Choiruddin, Jean-François Coeurjolly, and Frédérique Letué.
\newblock Convex and non-convex regularization methods for spatial point processes intensity estimation.
\newblock {\em Electronic Journal of Statistics}, 12(1), January 2018.

\bibitem{yue_variable_2015}
Yu~(Ryan) Yue and Ji~Meng Loh.
\newblock Variable selection for inhomogeneous spatial point process models.
\newblock {\em The Canadian Journal of Statistics / La Revue Canadienne de Statistique}, 43(2):288--305, 2015.
\newblock Publisher: [Statistical Society of Canada, Wiley].

\bibitem{ba_inference_2023}
Ismaïla Ba and Jean‐François Coeurjolly.
\newblock Inference for low‐ and high‐dimensional inhomogeneous {Gibbs} point processes.
\newblock {\em Scandinavian Journal of Statistics}, 50(3):993--1021, September 2023.

\bibitem{rajala_detecting_2018}
T.~Rajala, D.~J. Murrell, and S.~C. Olhede.
\newblock Detecting {Multivariate} {Interactions} in {Spatial} {Point} {Patterns} with {Gibbs} {Models} and {Variable} {Selection}.
\newblock {\em Journal of the Royal Statistical Society Series C: Applied Statistics}, 67(5):1237--1273, November 2018.

\bibitem{choiruddin_regularized_2020}
Achmad Choiruddin, Francisco Cuevas-Pacheco, Jean-François Coeurjolly, and Rasmus Waagepetersen.
\newblock Regularized estimation for highly multivariate log {Gaussian} {Cox} processes.
\newblock {\em Statistics and Computing}, 30(3):649--662, May 2020.

\bibitem{spychala_variable_2024}
Cécile Spychala, Clément Dombry, and Camelia Goga.
\newblock Variable selection methods for {Log}-{Gaussian} {Cox} processes: {A} case-study on accident data.
\newblock {\em Spatial Statistics}, 61:100831, June 2024.

\bibitem{zou_adaptive_2006}
Hui Zou.
\newblock The {Adaptive} {Lasso} and {Its} {Oracle} {Properties}.
\newblock {\em Journal of the American Statistical Association}, 101(476):1418--1429, December 2006.

\bibitem{coeurjolly_regularization_2023}
Jean-François Coeurjolly, Ismaïla Ba, and Achmad Choiruddin.
\newblock Regularization techniques for inhomogeneous (spatial) point processes intensity and conditional intensity estimation, May 2023.
\newblock arXiv:2305.13470 [math, stat].

\bibitem{bach_learning_2024}
F.~Bach.
\newblock {\em Learning {Theory} from {First} {Principles}}.
\newblock Adaptive {Computation} and {Machine} {Learning} series. MIT Press, 2024.

\bibitem{choiruddin_information_2021}
Achmad Choiruddin, Jean‐François Coeurjolly, and Rasmus Waagepetersen.
\newblock Information criteria for inhomogeneous spatial point processes.
\newblock {\em Australian \& New Zealand Journal of Statistics}, 63(1):119--143, March 2021.

\bibitem{guttorp_what_2023}
Peter Guttorp, Janine Illian, Joel Kostensalo, Mikko Kuronen, Mari Myllymäki, Aila Särkkä, and Thordis~L. Thorarinsdottir.
\newblock What you see is not what is there: {Mechanisms}, models, and methods for point pattern deviations, October 2023.
\newblock arXiv:2310.02292 [stat].

\bibitem{gillespie_deep_2024}
Lauren~E. Gillespie, Megan Ruffley, and Moises Exposito-Alonso.
\newblock Deep learning models map rapid plant species changes from citizen science and remote sensing data.
\newblock {\em Proceedings of the National Academy of Sciences}, 121(37):e2318296121, September 2024.

\bibitem{kuronen_point_2021}
Mikko Kuronen, Mari Myllymäki, Adam Loavenbruck, and Aila Särkkä.
\newblock Point process models for sweat gland activation observed with noise.
\newblock {\em Statistics in Medicine}, 40(8):2055--2072, April 2021.

\bibitem{briz-redon_dealing_2024}
Alvaro Briz-Redón.
\newblock Dealing with location uncertainty for modeling network-constrained lattice data.
\newblock {\em Spatial Statistics}, 59:100807, March 2024.

\bibitem{lund_models_2000}
Jens Lund and Mats Rudemo.
\newblock Models for point processes observed with noise.
\newblock {\em Biometrika}, 87(2):235--249, June 2000.

\bibitem{werner_loss-guided_2023}
Tino Werner.
\newblock Loss-guided stability selection.
\newblock {\em Advances in Data Analysis and Classification}, December 2023.

\bibitem{meinshausen_stability_2010}
Nicolai Meinshausen and Peter Bühlmann.
\newblock Stability {Selection}.
\newblock {\em Journal of the Royal Statistical Society Series B: Statistical Methodology}, 72(4):417--473, September 2010.

\bibitem{shah_variable_2013}
Rajen~D. Shah and Richard~J. Samworth.
\newblock Variable {Selection} with {Error} {Control}: {Another} {Look} at {Stability} {Selection}.
\newblock {\em Journal of the Royal Statistical Society Series B: Statistical Methodology}, 75(1):55--80, January 2013.

\bibitem{maddu_stability_2022}
Suryanarayana Maddu, Bevan~L. Cheeseman, Ivo~F. Sbalzarini, and Christian~L. Müller.
\newblock Stability selection enables robust learning of differential equations from limited noisy data.
\newblock {\em Proceedings of the Royal Society A: Mathematical, Physical and Engineering Sciences}, 478(2262):20210916, June 2022.

\bibitem{nogueira_stability_2018}
Sarah Nogueira, Konstantinos Sechidis, and Gavin Brown.
\newblock On the {Stability} of {Feature} {Selection} {Algorithms}.
\newblock {\em Journal of Machine Learning Research}, 18(174):1--54, 2018.

\bibitem{cronie_cross-validation-based_2024}
Ottmar Cronie, Mehdi Moradi, and Christophe A~N Biscio.
\newblock A cross-validation-based statistical theory for point processes.
\newblock {\em Biometrika}, 111(2):625--641, May 2024.

\bibitem{moller_modern_2007}
Jesper Møller and Rasmus~P. Waagepetersen.
\newblock Modern {Statistics} for {Spatial} {Point} {Processes}.
\newblock {\em Scandinavian Journal of Statistics}, 34(4):643--684, December 2007.

\bibitem{baddeley_spatial_2007}
Adrian Baddeley.
\newblock Spatial {Point} {Processes} and their {Applications}.
\newblock In Wolfgang Weil, editor, {\em Stochastic {Geometry}}, volume 1892 of {\em Lecture {Notes} in {Mathematics}}, pages 1--75. Springer Berlin Heidelberg, 2007.
\newblock Series Title: Lecture Notes in Mathematics.

\bibitem{coeurjolly_tutorial_2017}
Jean‐François Coeurjolly, Jesper Møller, and Rasmus Waagepetersen.
\newblock A {Tutorial} on {Palm} {Distributions} for {Spatial} {Point} {Processes}.
\newblock {\em International Statistical Review}, 85(3):404--420, December 2017.

\bibitem{lavancier_adaptive_2021}
Frédéric Lavancier, Arnaud Poinas, and Rasmus Waagepetersen.
\newblock Adaptive estimating function inference for nonstationary determinantal point processes.
\newblock {\em Scandinavian Journal of Statistics}, 48(1):87--107, March 2021.

\bibitem{moller_recent_2017}
Jesper Møller and Rasmus Waagepetersen.
\newblock Some {Recent} {Developments} in {Statistics} for {Spatial} {Point} {Patterns}.
\newblock {\em Annual Review of Statistics and Its Application}, 4(1):317--342, March 2017.

\bibitem{waagepetersen_estimating_2007}
Rasmus~Plenge Waagepetersen.
\newblock An {Estimating} {Function} {Approach} to {Inference} for {Inhomogeneous} {Neyman}–{Scott} {Processes}.
\newblock {\em Biometrics}, 63(1):252--258, March 2007.
\newblock Publisher: Oxford University Press (OUP).

\bibitem{guan_composite_2006}
Yongtao Guan.
\newblock A {Composite} {Likelihood} {Approach} in {Fitting} {Spatial} {Point} {Process} {Models}.
\newblock {\em Journal of the American Statistical Association}, 101(476):1502--1512, December 2006.

\bibitem{prokesova_two-step_2017}
Michaela Prokešová, Jiří Dvořák, and Eva B.~Vedel Jensen.
\newblock Two-step estimation procedures for inhomogeneous shot-noise {Cox} processes.
\newblock {\em Annals of the Institute of Statistical Mathematics}, 69(3):513--542, June 2017.

\bibitem{baddeley_non_2000}
A.~J. Baddeley, J.~Møller, and R.~Waagepetersen.
\newblock Non‐ and semi‐parametric estimation of interaction in inhomogeneous point patterns.
\newblock {\em Statistica Neerlandica}, 54(3):329--350, November 2000.

\bibitem{moller_statistical_2003}
Jesper Møller and Rasmus~Plenge Waagepetersen.
\newblock {\em Statistical {Inference} and {Simulation} for {Spatial} {Point} {Processes}}, volume 100 of {\em Monographs on {Statistics} and {Applied} {Probability}}.
\newblock Chapman and Hall/CRC, 1 edition, September 2003.

\bibitem{yang_fast_2020}
Yingzhen Yang and Jiahui Yu.
\newblock Fast {Proximal} {Gradient} {Descent} for {A} {Class} of {Non}-convex and {Non}-smooth {Sparse} {Learning} {Problems}.
\newblock In Ryan~P. Adams and Vibhav Gogate, editors, {\em Proceedings of {The} 35th {Uncertainty} in {Artificial} {Intelligence} {Conference}}, volume 115 of {\em Proceedings of {Machine} {Learning} {Research}}, pages 1253--1262. PMLR, July 2020.

\bibitem{parikh_proximal_2014}
Neal Parikh and Stephen Boyd.
\newblock Proximal {Algorithms}.
\newblock {\em Foundations and Trends® in Optimization}, 1(3):127--239, 2014.

\bibitem{barzilai_two-point_1988}
Jonathan Barzilai and Jonathan~M. Borwein.
\newblock Two-{Point} {Step} {Size} {Gradient} {Methods}.
\newblock {\em IMA Journal of Numerical Analysis}, 8(1):141--148, January 1988.

\bibitem{beck_fast_2009}
Amir Beck and Marc Teboulle.
\newblock A {Fast} {Iterative} {Shrinkage}-{Thresholding} {Algorithm} for {Linear} {Inverse} {Problems}.
\newblock {\em SIAM Journal on Imaging Sciences}, 2(1):183--202, January 2009.

\bibitem{choiruddin_adaptive_2023}
Achmad Choiruddin, Jean-François Coeurjolly, and Frédérique Letué.
\newblock Adaptive lasso and {Dantzig} selector for spatial point processes intensity estimation.
\newblock {\em Bernoulli}, 29(3), August 2023.

\bibitem{hui_tuning_2015}
Francis K.~C. Hui, David~I. Warton, and Scott~D. Foster.
\newblock Tuning {Parameter} {Selection} for the {Adaptive} {Lasso} {Using} {ERIC}.
\newblock {\em Journal of the American Statistical Association}, 110(509):262--269, January 2015.

\bibitem{bodinier_automated_2023}
Barbara Bodinier, Sarah Filippi, Therese~Haugdahl Nøst, Julien Chiquet, and Marc Chadeau-Hyam.
\newblock Automated calibration for stability selection in penalised regression and graphical models.
\newblock {\em Journal of the Royal Statistical Society Series C: Applied Statistics}, 72(5):1375--1393, December 2023.

\bibitem{hastie_elements_2009}
Trevor Hastie, Robert Tibshirani, and Jerome Friedman.
\newblock {\em The {Elements} of {Statistical} {Learning}}.
\newblock Springer {Series} in {Statistics}. Springer New York, New York, NY, 2009.

\bibitem{berman_approximating_1992}
Mark Berman and T.~Rolf Turner.
\newblock Approximating point process likelihoods with glim.
\newblock {\em Journal of the Royal Statistical Society Series C: Applied Statistics}, 41(1):31--38, 12 2018.

\bibitem{baddeley_spatial_2016}
Adrian Baddeley, Ege Rubak, and Rolf Turner.
\newblock {\em Spatial point patterns: methodology and applications with {R}}.
\newblock Chapman \& {Hall} / {CRC} {Interdisciplinary} {Statistics}. CRC Press, Taylor \& Francis Group, Boka Raton London New York, 2016.

\bibitem{baddeley_practical_2000}
Adrian Baddeley and Rolf Turner.
\newblock Practical {Maximum} {Pseudolikelihood} for {Spatial} {Point} {Patterns}: (with {Discussion}).
\newblock {\em Australian \& New Zealand Journal of Statistics}, 42(3):283--322, September 2000.

\bibitem{paszke_pytorch_2019}
Adam Paszke, Sam Gross, Francisco Massa, Adam Lerer, James Bradbury, Gregory Chanan, Trevor Killeen, Zeming Lin, Natalia Gimelshein, Luca Antiga, Alban Desmaison, Andreas K\"{o}pf, Edward Yang, Zach DeVito, Martin Raison, Alykhan Tejani, Sasank Chilamkurthy, Benoit Steiner, Lu~Fang, Junjie Bai, and Soumith Chintala.
\newblock {\em PyTorch: an imperative style, high-performance deep learning library}.
\newblock Curran Associates Inc., Red Hook, NY, USA, 2019.

\bibitem{condit_tropical_1998}
Richard Condit.
\newblock {\em Tropical {Forest} {Census} {Plots}}.
\newblock Springer Berlin Heidelberg, Berlin, Heidelberg, 1998.

\bibitem{hubbell_light-gap_1999}
S.~P. Hubbell, R.~B. Foster, S.~T. O'Brien, K.~E. Harms, R.~Condit, B.~Wechsler, S.~J. Wright, and S.~Loo De~Lao.
\newblock Light-{Gap} {Disturbances}, {Recruitment} {Limitation}, and {Tree} {Diversity} in a {Neotropical} {Forest}.
\newblock {\em Science}, 283(5401):554--557, January 1999.
\newblock Publisher: American Association for the Advancement of Science (AAAS).

\bibitem{virtanen_scipy_2020}
Pauli Virtanen, Ralf Gommers, Travis~E. Oliphant, and et~al.
\newblock {SciPy} 1.0: fundamental algorithms for scientific computing in {Python}.
\newblock {\em Nature Methods}, 17(3):261--272, March 2020.
\newblock Publisher: Springer Science and Business Media LLC.

\end{thebibliography}

\end{document}